\documentclass[12pt]{article}
\usepackage[utf8]{inputenc}
\usepackage[T1]{fontenc} % codage moderne des caractères sous Latex
\usepackage{lmodern}
\usepackage[english]{babel}           % style anglais

\usepackage{cancel}
\usepackage[pdftex]{graphicx}
\usepackage{epsfig}
\usepackage{graphicx}
\usepackage{comment}
\usepackage{latexsym}
\usepackage{hyperref}
\usepackage{amsmath}
\usepackage{mathrsfs}
\usepackage[usenames, dvipsnames]{color}
\usepackage{amsbsy}
\usepackage{amssymb}
\usepackage{amsthm}
\usepackage{amsfonts}
\usepackage{cite}
\usepackage{enumitem}
\usepackage{xcolor}
\usepackage{color}
\usepackage{mathtools}
\usepackage{caption} 
\usepackage{subcaption} %sous-figures 

\usepackage{float}

\usepackage{diagbox}

\usepackage[title]{appendix}

\usepackage{tabularx}

\renewcommand{\thefootnote}{\fnsymbol{footnote}}

\newcommand{\be}[0]{\begin{equation}}
\newcommand{\ee}[0]{\end{equation}}

\renewcommand{\thefootnote}{\fnsymbol{footnote}}

\newcommand{\Z}{\mathbb{Z}}

\renewcommand{\O}{{\cal O}}

\newcommand{\diag}{{\rm diag}\,}

\newcommand{\ie}{{\em i.e.} }

\newcommand{\via}{{\it via} }

\newcommand{\where}{\mbox{where}}

\renewcommand{\and}{\mbox{and}}
\newcommand{\for}{\mbox{for}}

\newcommand{\bm}{\boldmath} 
\newcommand{\V}{{\cal V}}
\newcommand{\W}{{\cal W}}
\newcommand{\F}{{\cal F}}
\newcommand{\N}{{\cal N}}
\newcommand{\J}{{\cal J}}
\newcommand{\I}{{\cal I}}

\renewcommand{\P}{{\cal P}}
\newcommand{\G}{{\cal G}}

\newcommand{\M}{{\cal M}}

\newcommand{\cR}{{\cal R}}

\newcommand{\Ms}{M_{\rm s}}

\newcommand{\nF}{n_{\rm F}}
\newcommand{\nB}{n_{\rm B}}

\newcommand{\half}{\frac{1}{2}}

\newcommand\dd{\text{d}}

\newcommand{\GP}{T^{4}/\mathbb{Z}_{2}}

\def\jac(#1,#2){%
\begin{bsmallmatrix}
#1\cr 
#2\cr
\end{bsmallmatrix}}

\catcode`\@=11
\def\marginnote#1{}
\newcount\hour
\newcount\minute
\newtoks\amorpm
\hour=\time\divide\hour by60 \minute=\time{\multiply\hour by60
\global\advance\minute by-\hour}
\edef\standardtime{{\ifnum\hour<12 \global\amorpm={am}%
        \else\global\amorpm={pm}\advance\hour by-12 \fi
        \ifnum\hour=0 \hour=12 \fi
        \number\hour:\ifnum\minute<10 0\fi\number\minute\the\amorpm}}
\edef\militarytime{\number\hour:\ifnum\minute<10 0\fi\number\minute}
\def\draftlabel#1{{\@bsphack\if@filesw {\let\thepage\relax
   \xdef\@gtempa{\write\@auxout{\string
      \newlabel{#1}{{\@currentlabel}{\thepage}}}}}\@gtempa
   \if@nobreak \ifvmode\nobreak\fi\fi\fi\@esphack}
        \gdef\@eqnlabel{#1}}
\def\@eqnlabel{}
\def\@vacuum{}
\def\draftmarginnote#1{\marginpar{\raggedright\scriptsize\tt#1}}
\def\draft{\oddsidemargin -.2truein
        \def\@oddfoot{\sl preliminary draft \hfil
        \rm\thepage\hfil\sl\today\quad\militarytime}
        \let\@evenfoot\@oddfoot \overfullrule 3pt
        \let\label=\draftlabel
        \let\marginnote=\draftmarginnote
   \def\@eqnnum{(\theequation)\rlap{\kern\marginparsep\tt\@eqnlabel}%
\global\let\@eqnlabel\@vacuum}  }
\def\thebibliography#1{
\vskip 0.5cm \centerline{\bf \Large References}
\list{
[\arabic{enumi}]}{\settowidth\labelwidth{[#1]}
\leftmargin\labelwidth
\advance\leftmargin\labelsep
\usecounter{enumi}}
\def\newblock{\hskip .11em plus .33em minus .07em}
\sloppy\clubpenalty4000\widowpenalty4000
\sfcode`\.=1000\relax}

\renewcommand{\theequation}{\arabic{section}.\arabic{equation}}
\renewcommand{\section}{\setcounter{equation}{0}\@startsection
{section}{1}{0mm}{-.\baselineskip}{0.5\baselineskip} {\normalfont\Large\bfseries}}
\renewcommand{\subsection}{\@startsection
{subsection}{2}{0mm}{-\baselineskip}{0.5\baselineskip} {\normalfont\large\bfseries}}
\renewcommand{\subsubsection}{\@startsection
{subsubsection}{3}{0mm}{-\baselineskip}{0.5\baselineskip}
{\normalfont\normalsize\slshape}}

\def\car(#1,#2,#3,#4,#5,#6,#7,#8){
\def\ArgI{{#1}}
\def\ArgII{{#2}}
\def\ArgIII{{#3}}
\def\ArgIV{{#4}}
\def\ArgV{{#5}}
\def\ArgVI{{#6}}
\def\ArgVII{{#7}}
\def\ArgVIII{{#8}}
}

\def\carRelay(#1,#2,#3){
\left| \ArgI_{4}\ArgII_{4}#1\ArgIII_{4}\ArgIV_{4}#2\ArgV_{4}\ArgVI_{4}#3\ArgVII_{4}\ArgVIII_{4}\right|^{2}
}

\def\caro(#1,#2,#3,#4,#5,#6,#7,#8){
\def\ArgI{{#1}}
\def\ArgII{{#2}}
\def\ArgIII{{#3}}
\def\ArgIV{{#4}}
\def\ArgV{{#5}}
\def\ArgVI{{#6}}
\def\ArgVII{{#7}}
\def\ArgVIII{{#8}}
}

\def\carRelayo(#1,#2,#3){
\left( \ArgI_{4}\ArgII_{4}#1\ArgIII_{4}\ArgIV_{4}#2\ArgV_{4}\ArgVI_{4}#3\ArgVII_{4}\ArgVIII_{4}\right)
}

\def\carm(#1,#2,#3,#4,#5,#6,#7,#8){
\def\ArgI{{#1}}
\def\ArgII{{#2}}
\def\ArgIII{{#3}}
\def\ArgIV{{#4}}
\def\ArgV{{#5}}
\def\ArgVI{{#6}}
\def\ArgVII{{#7}}
\def\ArgVIII{{#8}}
}

\def\carRelaym(#1,#2,#3){
\left( \hat{\ArgI}_{4}\hat{\ArgII}_{4}#1\hat{\ArgIII}_{4}\hat{\ArgIV}_{4}#2\hat{\ArgV}_{4}\hat{\ArgVI}_{4}#3\hat{\ArgVII}_{4}\hat{\ArgVIII}_{4}\right)
}

\def\carq(#1,#2,#3,#4,#5){
\left|#1_{4}#2_{4}#5#3_{4}#4_{4}\right|^{2}
}

\def\cars(#1,#2,#3,#4,#5){
(#1_{4}#2_{4}#5#3_{4}#4_{4})
}

\def\carsm(#1,#2,#3,#4,#5){
(\hat{#1}_{4}\hat{#2}_{4}#5\hat{#3}_{4}\hat{#4}_{4})
}

\def\carsbar(#1,#2,#3,#4,#5){
(\bar{#1}_{4}\bar{#2}_{4}#5\bar{#3}_{4}\bar{#4}_{4})
}

\def\LAMBDA(#1,#2){
\ifnum #2=0
\ifnum #1=0
\Lambda^{(2,2)}_{\vec{m},\vec{n}}
\else
\Lambda^{(2,2)}_{\vec{m}+\half,\vec{n}}
\fi
\else
\ifnum #1=0
\ifnum #2=0
\Lambda_{\vec{m},\vec{n}}
\else
\Lambda^{(2,2)}_{\vec{m},\vec{n}+\vec{a_{\text{S}}}}
\fi
\fi
\fi
\ifnum #1=1
\ifnum #2=1
\Lambda^{(2,2)}_{\vec{m}+\half,\vec{n}+\vec{a_{\text{S}}}}
\fi
\fi
}

\def\LAMBDA(#1,#2){
\frac{\Lambda^{(2,2)}_{#1,#2}}{\left|\eta^{4}\right|^{2}}
}

\def\LAMBDAO(#1,#2){
\frac{\Lambda^{(2,2)}_{#1,#2}}{\eta^{4}}
}

\def\LAMBDAM(#1,#2){
\frac{\Lambda^{(2,2)}_{#1,#2}}{\hat{\eta}^{4}}
}

\makeatletter
\newcommand{\vast}{\bBigg@{3.5}}
\makeatother

\begin{document}

%%%%%

\begin{titlepage}
\begin{flushright}
CPHT-PC-013.032020, May 2020
\vspace{1.5cm}
\end{flushright}
\begin{centering}
{\bm\bf \Large Moduli stability in type I string orbifold models\footnote{Based on a talk given at the ``Humboldt Kolleg Frontiers in Physics: From the Electroweak to the Planck Scales,'' 15--19 September 2019, Corfu, Greece.}}

\vspace{7mm}

 {\bf Thibaut Coudarchet and Herv\'e Partouche}

 \vspace{4mm}

{CPHT, CNRS, Ecole polytechnique, IP Paris, \\F-91128 Palaiseau, France \\ 
\textit{thibaut.coudarchet@polytechnique.edu}\\\textit{herve.partouche@polytechnique.edu}}

\end{centering}
\vspace{0.1cm}
$~$\\
\centerline{\bf\Large Abstract}\\
\vspace{-0.7cm}

\begin{quote}

\hspace{.6cm} 
We analyze the stability of the moduli at the quantum  level in an open-string model realizing the  $\N=2\to \N=0$ spontaneous breaking of supersymmetry in four-dimensional Minkowski spacetime. In the region of moduli space where the supersymmetry breaking scale is lower than the other scales, we identify vanishing minima of the one-loop effective potential, up to exponentially small corrections. In these backgrounds, the spectrum satisfies Bose-Fermi degeneracy at the massless level.

\end{quote}

\end{titlepage}
\newpage
\setcounter{footnote}{0}
\renewcommand{\thefootnote}{\arabic{footnote}}
 \setlength{\baselineskip}{.7cm} \setlength{\parskip}{.2cm}

\setcounter{section}{0}

%%%%%%%%%%%%%%%%%%%%%%%%%%%%%%%%%%%%%%%%%%%%%%%%%%%%%%%%%%%%%%%%%%%

\section{Introduction}

In the present work, we address the question of tachyonic instabilities at the quantum level in type I string orbifold models showing an $\N=2\to \N=0$  spontaneous breaking of supersymmetry in four dimensions~\cite{Paper}. This is done by developing a global geometric picture of the potential, first described in \cite{PreviousPaper,Partouche:2019pgv} in the case of $\N=4\to \N=0$ models. The general route of breaking supersymmetry at the classical level in flat space, and analyzing the induced effective potential,  was advocated in Refs~\cite{Itoyama:1986ei,Abel:2015oxa,SNS1,SNS2,FR,Abel:2017rch,Abel:2017vos,CatelinJullien:2007hw,Bourliot:2009na,CFP,Borunda:2002ra}. Moreover, the question of stability was addressed in the heterotic string framework in~\cite{SNS1,SNS2,APP,sta3,sta1,sta2,CoudarchetPartouche,Itoyama:2020ifw}. The supersymmetry breaking is implemented by a string version   of the Scherk-Schwarz mechanism~\cite{SS}, developed in the open string context in Refs~\cite{openSS2,openSS3,openSS4,openSS5,openSS6}. If the internal space involved in the mechanism is a circle of radius $R_5$, the supersymmetry breaking scale $M$ defined as the gravitini mass is 
\begin{equation}
M=\frac{\Ms}{2R_{5}}~,
\end{equation}
where $\Ms$ is the string scale. 
When the radius is sufficiently large, the dominant contribution to the potential at one loop is determined by the massless states and their Kaluza-Klein (KK) towers along the Scherk-Schwarz direction. More precisely, if no mass scale between $0$ and $M$ is present in the model, the one-loop effective potential around a critical background reads in $d$ dimensions \cite{PreviousPaper,Itoyama:1986ei,Abel:2015oxa,SNS1,SNS2,FR,Abel:2017rch,Abel:2017vos,CatelinJullien:2007hw,Bourliot:2009na,CFP,sta3,sta1,sta2,CoudarchetPartouche}
\begin{equation}
\label{rule_of_thumb}
\V\simeq(\nF-\nB)\xi_{d}M^{d}~,
\end{equation}
up to exponentially small terms. In this expression, $\nF$ and $\nB$  count the numbers of massless fermionic and bosonic degrees of freedom, while $\xi_d$ is a positive dressing arising from the KK towers. Moreover, any potential tree-level instability occurring when $M$ is of the order of $M_{\text{s}}$~\cite{PV1,PV2}, which are related to the Hagedorn transition, are avoided.

From the above equation, we see that backgrounds which have more massless fermions than bosons yield a positive potential. However, the Scherck-Schwarz mechanism implemented alone provides a mass shift to the fermions of the theory and consequently leads generically to a negative vacuum energy. Uplifting the potential thus requires additional ingredients. For instance, the introduction of open string Wilson lines (WL's) allows to increase the value of the potential by counterbalancing the Scherck-Schwarz shift,  in order to keep fermions massless and to  induce masses for bosons~\cite{PreviousPaper}. In this paper, we explicitly show the existence of models that have an exponentially small potential at one loop, \ie with $\nF-\nB=0$, as such models could constitute the groundwork needed to generate a small cosmological constant. The idea is that an exponentially suppressed one-loop potential may conspire with higher-loops effects to stabilise $M$ and the dilaton, and eventually yield a cosmological term smaller than in generic models.

The question of the sign or vanishing of the dominant contribution~(\ref{rule_of_thumb})  of the potential must be supplemented by the inspection of the stability of the background with respect to the various WL's and other moduli. The stability of the WL's is dictated by the signs of the second-order terms in the Taylor expansion of the potential at a given background. Non-negativity of all squared masses is required to obtain (marginally) stable configurations. The WL masses at one loop are given by the difference between the Dynkin indices of the representations in which the massless bosons are organized and the Dynkin indices of the massless fermionic representations \cite{SNS1,SNS2,CoudarchetPartouche}. Thus, non-negativity of the squared masses requires the contributions of the massless bosons to dominate those of the massless fermions, while for the potential to be positive, $\nF$ should be greater than $\nB$. This explains why  finding stable vacua with a positive (in which case $M$ undergoes a runaway) or exponentially suppressed potential is not a trivial task.

In Sect.~\ref{BSGP_deformation}, we detail the construction of our $\N=2\to \N=0$ orbifold setup in type I string, starting from the original supersymmetric Bianchi-Sagnotti-Gimon-Polchinski (BSGP) model \cite{Bianchi-Sagnotti,GimonPolchinski,GimonPolchinski2} compactified down to four dimensions and  taking into account marginal deformations. Then, the supersymmetry breaking is implemented with the Scherk-Schwarz mechanism in a direction orthogonal to the orbifold action. In Sect.~\ref{sc}, we compute the mass correction at one loop of the WL's, and discuss the (marginal) stability of the closed-string moduli. Finally, in Sect.~\ref{models}, we scan by computer all points in moduli space where the one-loop effective potential is extremal with respect to the WL's. We list all backgrounds tachyon free at one loop that show an exponentially small potential $(\nF-\nB=0)$, or a positive potential $(\nF-\nB>0)$ with runaway behavior of $M$. Further details beyond the results presented here can be found in \cite{Paper}.

%%%%%%%%%%%%%%%%%%%%%%%%%%%%%%%%%%%%%%

\section{\bm${\N=2\to \N=0}$ open string model}
\label{BSGP_deformation}

\subsection{Construction of the model}
\label{21}

The starting point is the supersymmetric Bianchi-Sagnotti-Gimon-Polchinski (BSGP) model \cite{Bianchi-Sagnotti,GimonPolchinski,GimonPolchinski2}. It consists in the orientifold projection of the type IIB superstring compactified on $T^4/\Z_2$. The $\mathbb{Z}_{2}$ generator $g$ acts as  $(X^{6},X^{7},X^{8},X^{9})\to -(X^{6},X^{7},X^{8},X^{9})$ on the coordinates of $T^{4}$. In this model, the Ramond-Ramond (RR) tadpole cancellation imposes the presence of $32$ D9-branes and $32$ D5-branes orthogonal to $T^{4}$. They cancel the charges of the O9-plane and O5-planes that are respectively the loci of fixed points of the orientifold generator $\Omega$ and of the combination $\Omega g$. We further compactify down to four dimensions by introducing a torus $T^{2}$, which leads to the background 
\begin{equation}
\mathbb{R}^{1,3}\times T^{2}\times \GP~.
\end{equation}
$T^{2}$ is assumed to wrap directions $4$ and $5$. The metric of both tori is denoted $G_{\I\J}$, $\I,\J=4,\dots 9$, and we define two sets of non-calligraphic indices to refer to the $T^{2}$ directions only, or to the $T^{4}$ ones only, $I'=4,5$ and $I=6,\dots,9$. 

Consistency conditions require the algebra of Chan-Patton factors to correspond to unitary or symplectic gauge groups instead of orthogonal ones \cite{Bianchi-Sagnotti,GimonPolchinski}. The original model showing a $U(16)\times U(16)$ gauge group and $\N=2$ supersymmetry in four dimensions can be generalized by introducing all sorts of marginal deformations. First, arbitrary positions of the D5-branes along $T^{4/}\Z_2$ can be turned on. Second, WL's along $T^{2}$ can be introduced for the gauge group associated with the D5-branes, and eventually Wilson lines along all of the six internal directions can be switched on for the gauge group associated with the D9-branes. Moduli in the Neuman (N)-Dirichlet (D) sector may also exist, and will be analysed in a subsequent work \cite{wip}. In the closed string sector, beside the internal metric $G_{\I\J}$ and the dilaton in the Neuveu-Schwarz-Neuveu-Schwarz (NS-NS) sector, there are moduli in the Ramond-Ramond (RR) sector associated with the two-form $C_{\I\J}$. In the twisted sector, there are also 16 quaternionic moduli localized at each of the 16 fixed points of $T^4/\Z_2$.

\begin{figure}
\captionsetup[subfigure]{position=t}
\begin{center}
\begin{subfigure}[t]{0.48\textwidth}
\begin{center}
\includegraphics [scale=0.55]{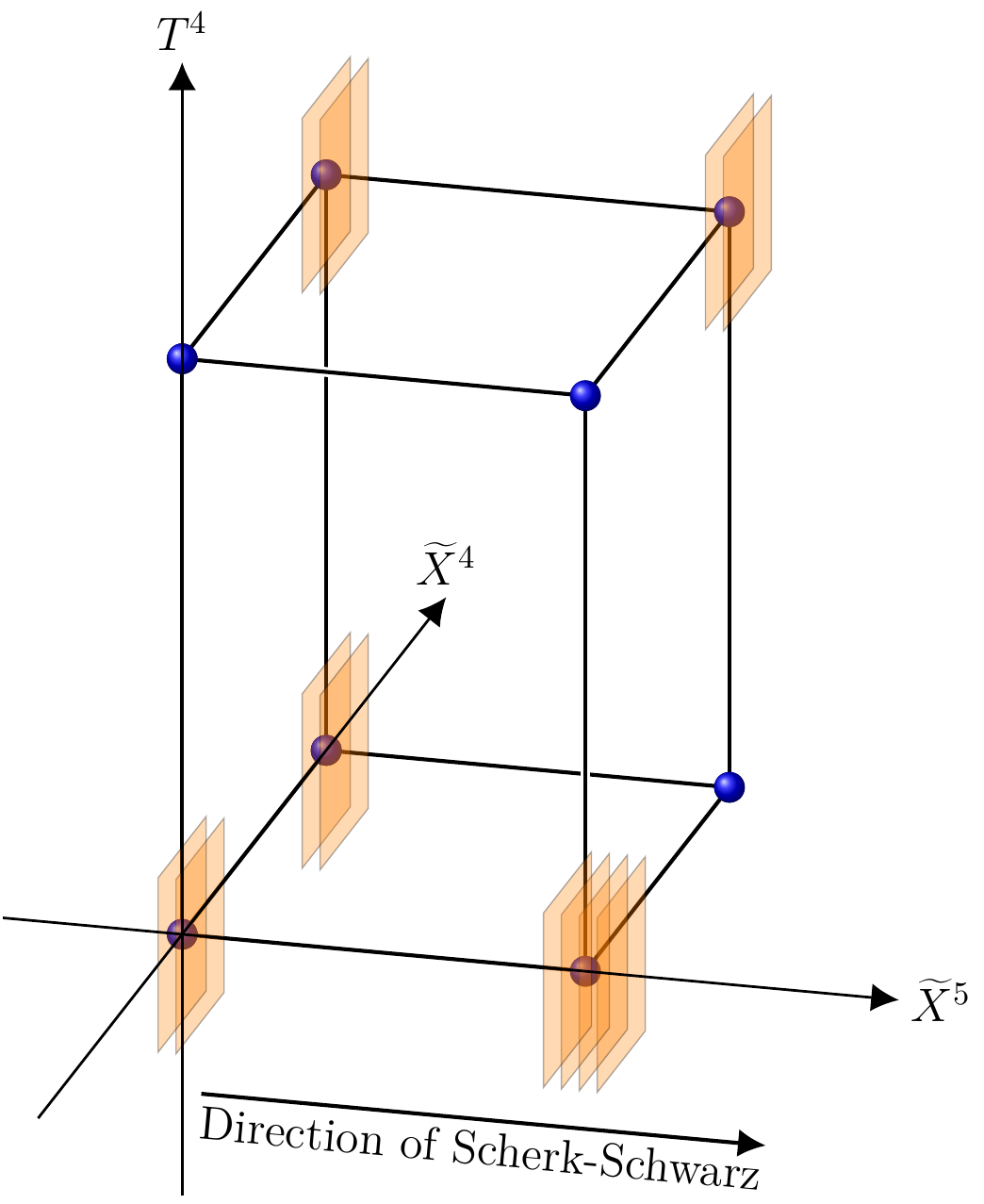}
\end{center}
\caption{\footnotesize A configuration of D3-branes associated with the D5-branes of the initial type~I theory, once $T^{2}$ is T-dualized. In this example, the D3-branes sit on O3-planes.}
\label{D5}
\end{subfigure}
\quad
\begin{subfigure}[t]{0.48\textwidth}
\begin{center}
\includegraphics [scale=0.55]{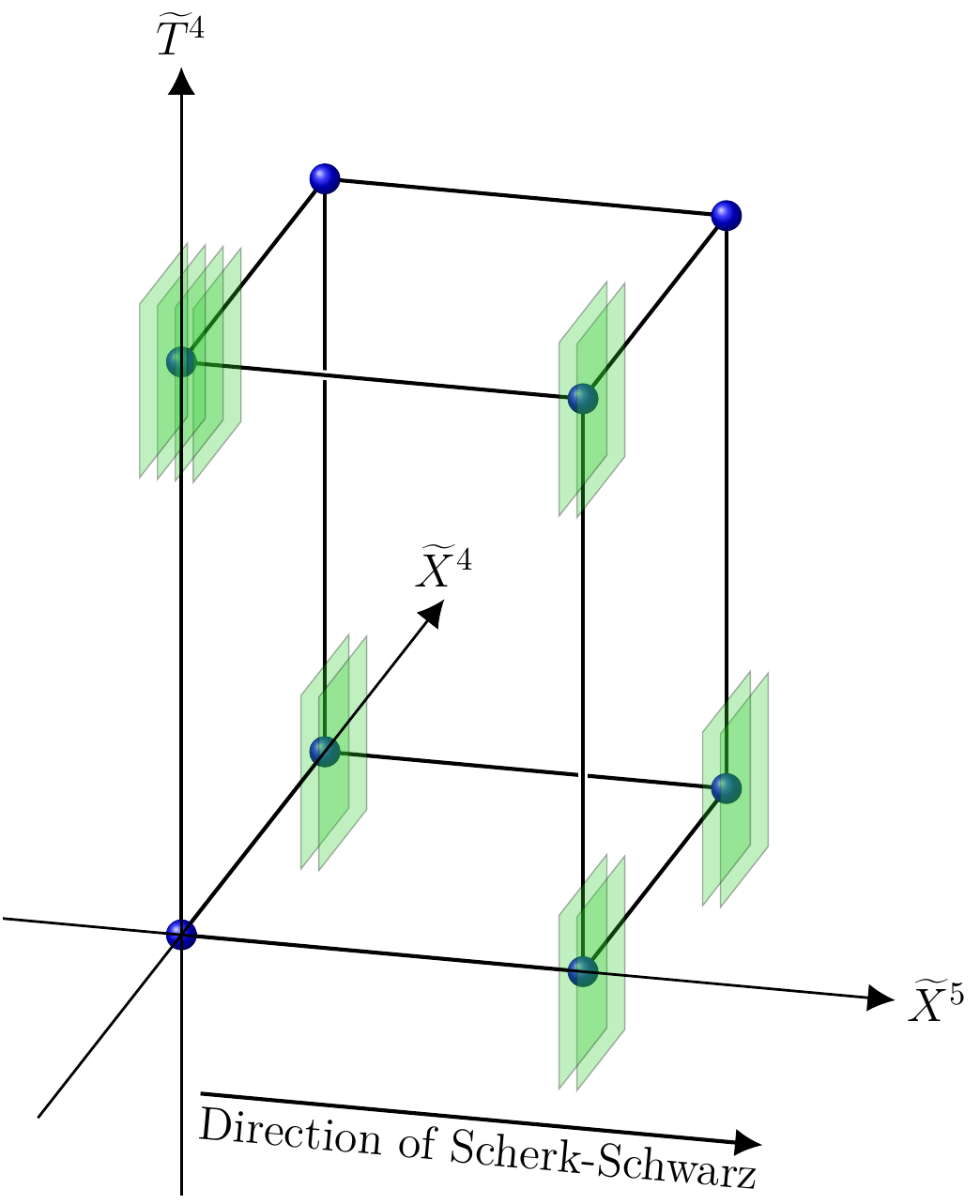}
\end{center}
\caption{\footnotesize A configuration of D3-branes associated with the D9-branes of the initial type~I theory, once both $T^{2}$ and $T^4/\Z_2$ are T-dualized. In this example, the D3-branes sit on O3-planes.}
\label{D9}
\end{subfigure}
\begin{subfigure}[t]{0.48\textwidth}
\begin{center}
\includegraphics [scale=0.55]{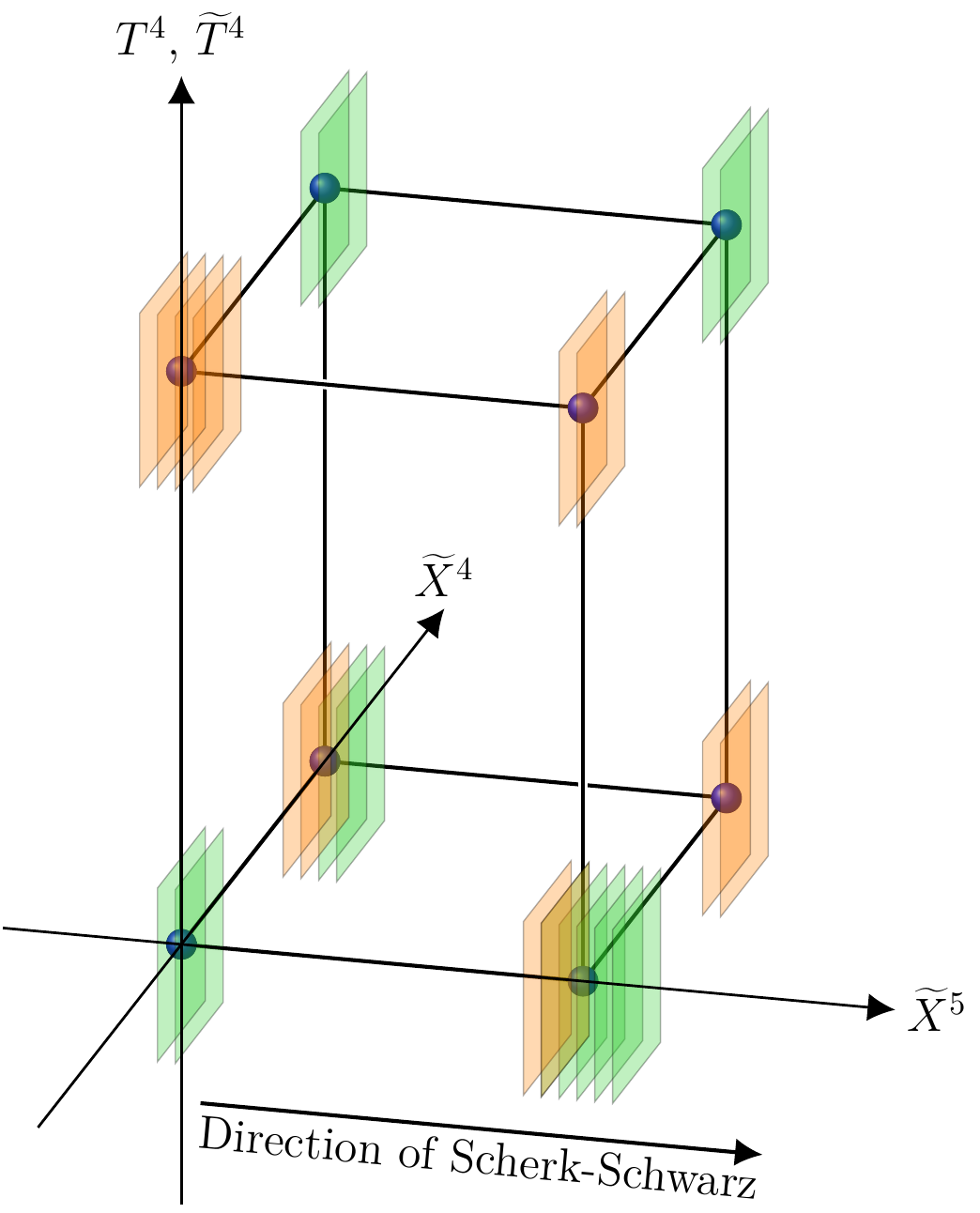}
\end{center}
\caption{\footnotesize Superposition of pictures (a) and (b). D3-branes associated with the  D5-branes (D9-branes) of the initial type~I theory are shown in orange (green).}
\label{D5D9}
\end{subfigure}
\quad
\begin{subfigure}[t]{0.48\textwidth}
\begin{center}
\includegraphics [scale=0.5]{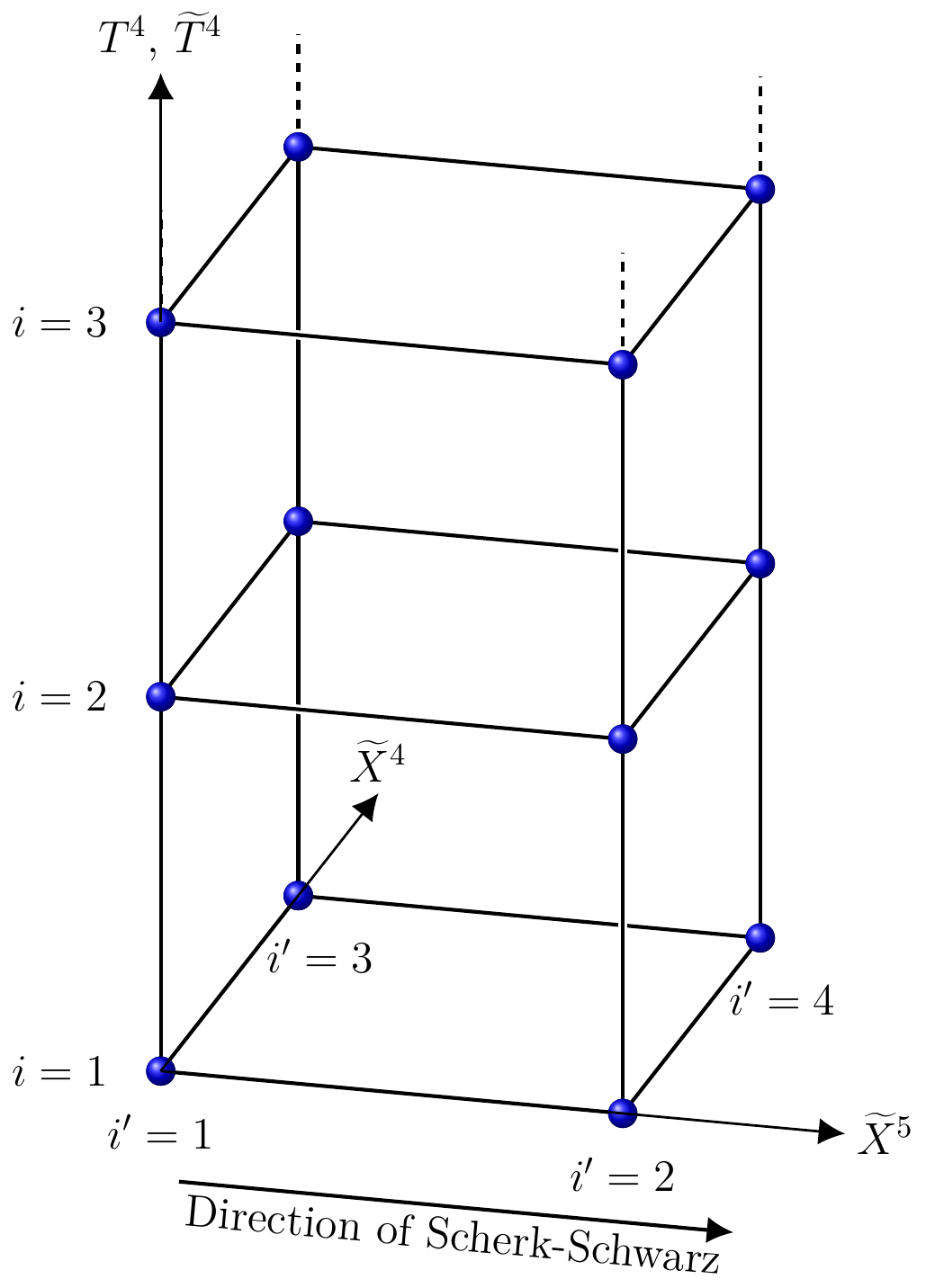}
\end{center}
\caption{\footnotesize Labelling of the $\tilde T^2$ fixed points $i'=1,2,3,4$, and schematic labelling of the $T^4/\Z_2$ or $\tilde T^4/\Z_2$ fixed points $i=1,\dots, 16$. Odd $i'$ correspond to  points located at $\tilde X^5=0$, while even $i'$ are associated with points at $\tilde X^5=\pi$, where $\tilde X^5$ is the coordinate T-dual to the direction along which the Scherk-Schwarz mechanism is implemented.}
\label{corners}
\end{subfigure}
\caption{\footnotesize Geometric T-dual description of the moduli arising from the NN and DD sectors of the orientifold theory. 
%The first two pictures show the geometric interpretation  of the deformations in two different theories. The third picture is the combination of those two into one abusive but useful representation. The last drawing shows our corner labelling of the internal box.
}
\label{D5D9D5D9}
\end{center}
\end{figure}

At generic brane positions or open string WL's, the original gauge group is spontaneously broken. Because of the orbifold and orientifold actions, the D5-branes can only move by packets of four along $T^{4}/\Z_2$. If $2n$ branes are at a fixed point, this creates a gauge factor $U(n)$ that can break into $U(n-2k)\times USp(2k)$ if $2k$ branes move away from the fixed point together with their $2k$ mirror branes with opposite coordinates in $T^4$. This implies the moduli space to be split into different disconnected components corresponding to different numbers of branes with rigid positions in $T^4/\Z_2$. Indeed, when there are $2n'+2$ branes at a fixed point, two of them cannot move in $T^4/\Z_2$. There are therefore a maximum of 8 independent positions in $T^4/\Z_2$. The WL's along $T^2$ of the gauge group generated by the D5-branes can also be given a geometric interpretation. This is done by T-dualizing the torus $T^{2}$ into $\tilde{T}^2$, which implies the WL's to become positions. Overall, the D5-branes  become D3-branes orthogonal to all internal directions. The coordinates of $\tilde T^{2}$ are denoted $\tilde{X}^{4}$ and $\tilde{X}^{5}$. 
Along  $T^{4}$, the orbifold identifies points that are mirrors to each other, while along $\tilde{T}^{2}$, the orientifold also creates a mirror setup \cite{review-3} (we then write $\tilde{T}^2/I_{45}$, where $I_{45}$ is the inversion $(\tilde{X}^{4},\tilde{X}^{5})\to -(\tilde X^4,\tilde X^5)$), implying the D3-branes to move by packets of two. There are therefore $16$ independent positions along $\tilde T^2$. 
  In this T-dual picture, all the internal space can be represented as a six-dimensional "box" $\tilde T^2/I_{45}\times T^4/\Z_2$, with an O3-plane at each of the 64 fixed points, and along which D3-branes can move, as depicted in Fig.~\ref{D5}.

On the other hand, in order to convert the WL's associated with the D9-branes into positions, we need to T-dualize all six internal directions. Denoting $\tilde T^4$ the torus dual to $T^{4}$, the D9-branes are converted into D3-branes orthogonal to all internal directions $\tilde X^4,\dots, \tilde X^9$, and there are 64 O3-planes distributed on the 64 fixed points of $\tilde T^2/I_{45}\times \tilde T^4/\Z_2$. This is shown schematically in Fig.~\ref{D9}. Notice that the descriptions in which the deformations of the D9-brane and D5-brane sectors are geometrical are distinct. However, for the sake of simplicity, we will represent things on the same picture (see Fig.~\ref{D5D9}). Even if it is abusive, this turns out to be very useful to understand and manipulate various models. For this reason, we will talk about brane positions or WL's interchangeably. It is also  understood that all positions refer to the appropriate T-dual descriptions. In fact, before the orbifold action is implemented, the Wilson line matrices associated with the D9-branes live in the Cartan subgroup of $SO(32)$,
\begin{align}
\begin{split}
\W_{\I}^{\text{D9}}&=\diag\left(e^{2i\pi a_{\alpha}^{\I}},\alpha=1,\dots,32\right)\;\;\;\for\;\;\; \I=4,\dots,9\\
&=\diag\left(e^{2i\pi a_{1}^{\I}},e^{-2i\pi a_{1}^{\I}},e^{2i\pi a_{2}^{\I}},e^{-2i\pi a_{2}^{\I}},\dots,e^{2i\pi a_{16}^{\I}},e^{-2i\pi a_{16}^{\I}}\right)~.
\end{split}
\end{align}
This parametrisation remains valid in the orbifold model for the matrices $\W^{\rm D9}_{I'}, I'=4,5$, associated with $T^{2}$. However, the number of moduli fields associated with the $T^{4}/\Z_2$ directions $I=6,7,8,9$ is reduced, and takes at most the following form, 
\begin{align}
\begin{split}
&\W_{I'}^{\text{D9}}=\diag\left(e^{2i\pi a_{1}^{I'}},e^{-2i\pi a_{1}^{I'}},e^{2i\pi a_{2}^{I'}},e^{-2i\pi a_{2}^{I'}},\dots,e^{2i\pi a_{16}^{I'}},e^{-2i\pi a_{16}^{I'}}\right)~,\\
&\W_{I}^{\text{D9}}=\diag\left(e^{2i\pi a_{1}^{I}},e^{-2i\pi a_{1}^{I}},\dots,e^{2i\pi a_{8}^{I}},e^{-2i\pi a_{8}^{I}},e^{2i\pi a_{1}^{I}},e^{-2i\pi a_{1}^{I}},\dots,e^{2i\pi a_{8}^{I}},e^{-2i\pi a_{8}^{I}}\right)~.
\end{split}
\end{align}
In the T-dual picture, the positions of the D3-branes along $\tilde{X}^{\I}$ are $2\pi a_{\alpha}^{\I},\ \I=4,\dots,9$. On the other hand, the positions in $\tilde T^2/I_{45}\times T^4/\Z_2$ of the D3-branes T-dual to the D5-branes are denoted $2\pi b_{\alpha}^{\I}$, $\I=4,\dots, 9$, which can also be collected in WL matrices $\W_{\I}^{\text{D5}}$. Note that according to what was said earlier, the WL matrices $\W_{I}^{\text{D9}}$ and $\W_{I}^{\text{D5}}$, $I=6, \dots, 9$, contain less than 8 degrees of freedom when pairs of D3-branes have rigid positions at fixed points of  $\tilde T^4/\Z_2$ or $T^4/\Z_2$.

Configurations where all D3-branes are located at the corners of the internal box (\ie sitting on the O3-planes of the appropriate T-dual descriptions) are of particular interest since they guarantee the potential to be critical with respect to the marginal deformations, as will be seen in Sect.~\ref{mT1}. In this case, it is convenient to introduce a specific labelling for the internal corners. The latter are designated by two indices $ii'$, where $i\in\{1,\dots,16\}$ labels the fixed points of $T^{4}/\Z_2$ (or $\tilde{T}^{4}/\Z_2$),  and $i'\in\{1,\dots,4\}$ labels those of $\tilde{T}^{2}/I_{45}$,
%For later convenience, when we introduce the Sherck-Schwarz mechanism along the direction $X^5$, we choose the corners $i'=1,3$ and $i'=2,4$  to be respectively at $\tilde X^5=0$ and $\tilde X^5=\pi$. 
as shown in Fig.~\ref{corners}. At a given corner $ii'$, we call $N_{ii'}$ the number of D3-branes T-dual to D9-branes and $D_{ii'}$ the number of D3-branes T-dual to D5-branes. In this setup, all Wilson lines $a_\alpha^\I$, $b_\alpha^\I$, $\I=4,\dots,9$,  take values equal to 0 or $\half$. In terms of vectors with six components, they take values $\vec{a}_{ii'}$, where $2\pi\vec{a}_{ii'}$ is the position of the corner~$ii'$.  These vectors can be decomposed along  $\tilde{T}^2/I_{45}$ and $T^4/\Z_2$ (or $\tilde T^4/\Z_2$) as $\vec{a}_{ii'}=(\vec{a}_{i'},\vec{a}_i)$.

In these supersymmetric configurations, the number of branes $N_{ii'}$, $D_{ii'}$ and their counterparts $R^{\text{N}}_{ii'}$ and $R^{\text{D}}_{ii'}$ \cite{PradisiSagnotti,review-1,review-2} under the orbifold action are parametrised as 
\begin{equation}
\label{unitaryparam}
N_{ii'}=n_{ii'}+\bar{n}_{ii'}~,\qquad D_{ii'}=d_{ii'}+\bar{d}_{ii'}~,\qquad R^{\text{N}}_{ii'}=i(n_{ii'}-\bar{n}_{ii'})~,\qquad R^{\text{D}}_{ii'}=i(d_{ii'}-\bar{d}_{ii'})~,
\end{equation}
with $n_{ii'}=\bar{n}_{ii'}$ and $d_{ii'}=\bar{d}_{ii'}$. The tadpole cancellation condition implies
\begin{equation}
\sum_{i,i'}N_{ii'}=32~\Longleftrightarrow~\sum_{i,i'}n_{ii'}=16~,\qquad\qquad \sum_{i,i'}D_{ii'}=32~\Longleftrightarrow~\sum_{i,i'}d_{ii'}=16~,
\end{equation}
leading to the open string gauge group
\begin{equation}
\G_{\text{open}}=\prod_{i,i'}U(n_{ii'})\times U(d_{ii'})~.
\end{equation}

The final step is to implement the spontaneous breaking of supersymmetry \via a stringy version~\cite{openSS2,openSS3,openSS4,openSS5,openSS6} of the Scherk-Schwarz mechanism~\cite{SS}. This is done by implementing a free orbifold action on the fifth direction, $X^{5}\rightarrow X^{5}+\pi$, coupled to the operator $(-1)^{F}$, where $F$ is the spacetime fermion number. As a result, the gravitini acquire a mass 
\begin{equation}
\label{breakingscale}
M~=~\frac{\sqrt{G^{55}}}{2}\,\Ms~,
\end{equation}
which is therefore the scale of $\N=2\to \N=0$ spontaneous breaking of supersymmetry. 
$M$ itself is one of the marginal deformations, provided it is less than the critical value of order of the string scale $\Ms$, at which a tree-level tachyonic instability arises~\cite{PV1,PV2}. In the $T^2$ lattice, the Scherk-Schwarz mechanism translates into a shift $F\vec a'_S$ of the KK integer momentum $\vec{m}'=(m_4,m_5)$, where $\vec a'_S=(0,\half)$. As described above, when the WL deformations are discrete (the D3-branes sit on the O3-planes of the appropriate six-dimensional boxes), their values along $T^2$ take values equal to some $\vec a_{i'}$, $i'=1,\dots,4$. This has an important consequence on the light spectrum since KK modes in the open string sector are massless when 
\be
\vec m'+F\,\vec a'_S+\vec a_{i'}-\vec a_{j'} ~=~ \vec 0
\label{T2la}
\ee
vanishes, and that this equation admits solutions both for bosons $(F=0)$ and fermions ($F=1$). This will be detailed in  the next subsection and to this end, it is relevant to specify further the labelling of the $\tilde T^2/I_{45}$ fixed points.   We will denote by $i'=1,3$ those located at the origin of the T-dual Scherk-Schwarz direction, $\tilde X^5=0$, and by $i'=2,4$ those at $\tilde X^5=\pi$ (see Fig.~\ref{corners}).

The model constructed so far must satisfy additional requirements to remain valid at the non-perturbative level~\cite{GimonPolchinski2}. To state these additional constraints, let us first consider the BSGP model in six dimensions. The disconnected parts of the moduli space are characterized by the even number $\cR=0,2,\dots, 16$ of pairs of D5-branes mirror to each other with respect to $\Omega$, and that have rigid positions at distinct fixed points of $T^4/\Z_2$. To be consistent non-perturbatively, a model must have $\cR=0$, 8 or 16. 
When $\cR=8$, the mirror pairs must sit on the 8 corners of one of the hyperplanes $X^I=0$ or $\pi$, $I=6,\dots,9$. Similarly, the number of mirror pairs of D5-branes T-dual to the D9-branes with rigid positions in $\tilde T^4/\Z_2$ must be $\tilde \cR=0$, 8 or 16.  
Hence, there are only $3\times 3$ fully consistent components in the moduli space, which can be further reduced to 6 by T-duality:\footnote{They can  be connected to each other by deforming $T^4/\Z_2$ into smooth $K3$ manifolds~\cite{GimonPolchinski2}.}
\be
(\cR,\tilde \cR)=(0,0)~,~~ (0,8)~,~~ (0,16)~,~~ (8,8)~,~~ (8,16)~,~~ (16,16)~. 
\ee
Compactifying down to four dimensions and T-dualizing $T^2$, there are no additional constraints on the distribution of the D3-branes. The latter, including the $2\cR+2\tilde \cR$ ones with rigid positions in $T^4/\Z_2$ or $\tilde T^4/\Z_2$, can move along the directions of $\tilde T^2/I_{45}$.

%%%%%%%%%%%%%%%%

\subsection{Massless spectrum}
\label{massless_spectrum_section}

\begin{figure}[h!]
\captionsetup[subfigure]{position=t}
\begin{center}
\begin{subfigure}[t]{0.48\textwidth}
\begin{center}
\includegraphics [trim=0cm 16cm 10cm 0cm,clip,scale=0.55]{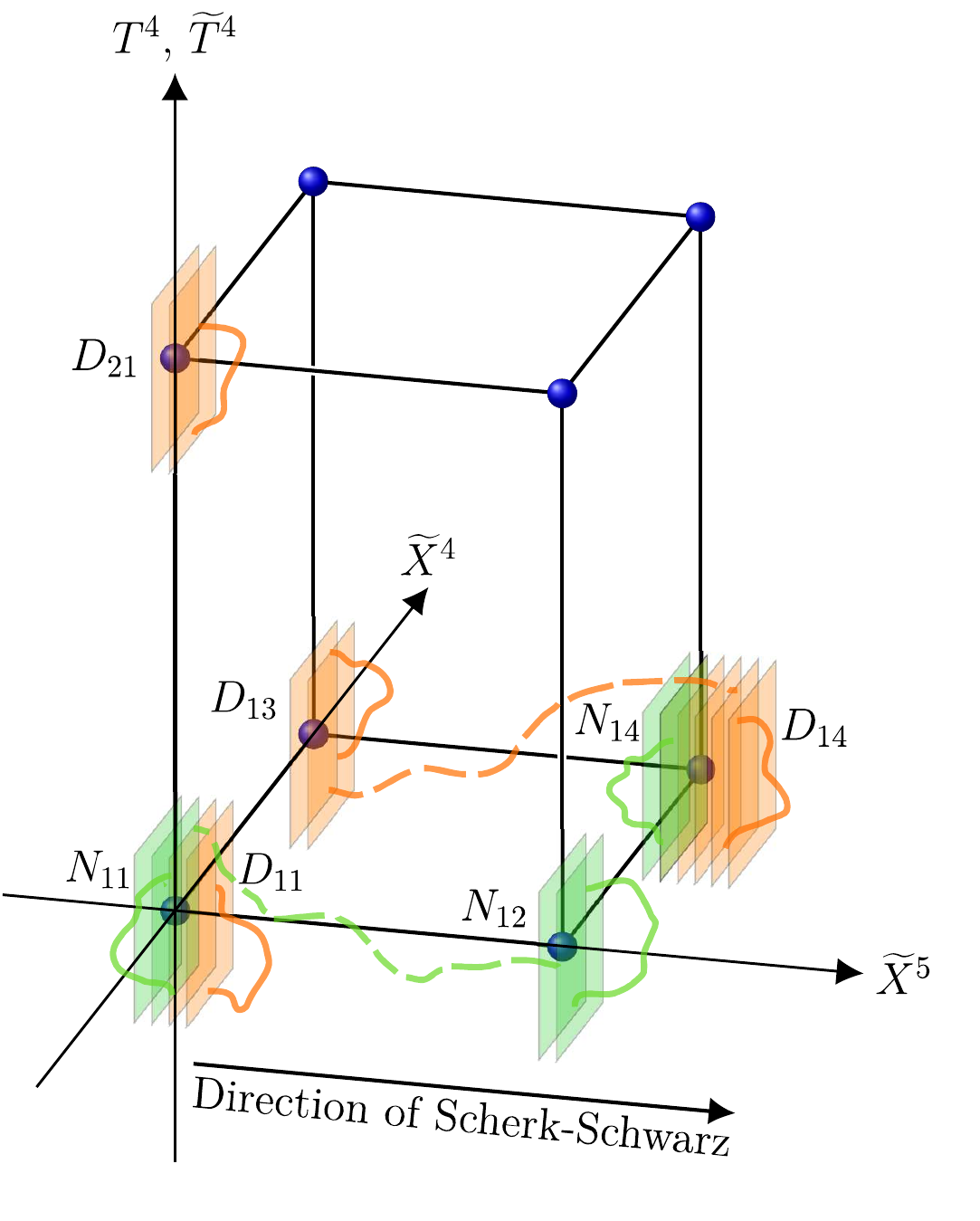}
\end{center}
\caption{NN and DD states are massless bosons when they correspond in the D3-brane picture to strings with both ends attached to the same stack of branes (solid strings). They are massless fermions when they correspond to strings stretched between corners of the six-dimensional box that are adjacent along the \mbox{T-dual} Scherk-Schwarz direction (dashed strings).}
\label{massless_NNDD}
\end{subfigure}
\quad
\begin{subfigure}[t]{0.48\textwidth}
\begin{center}
\includegraphics [trim=0cm 16cm 10cm 0cm,clip,scale=0.55]{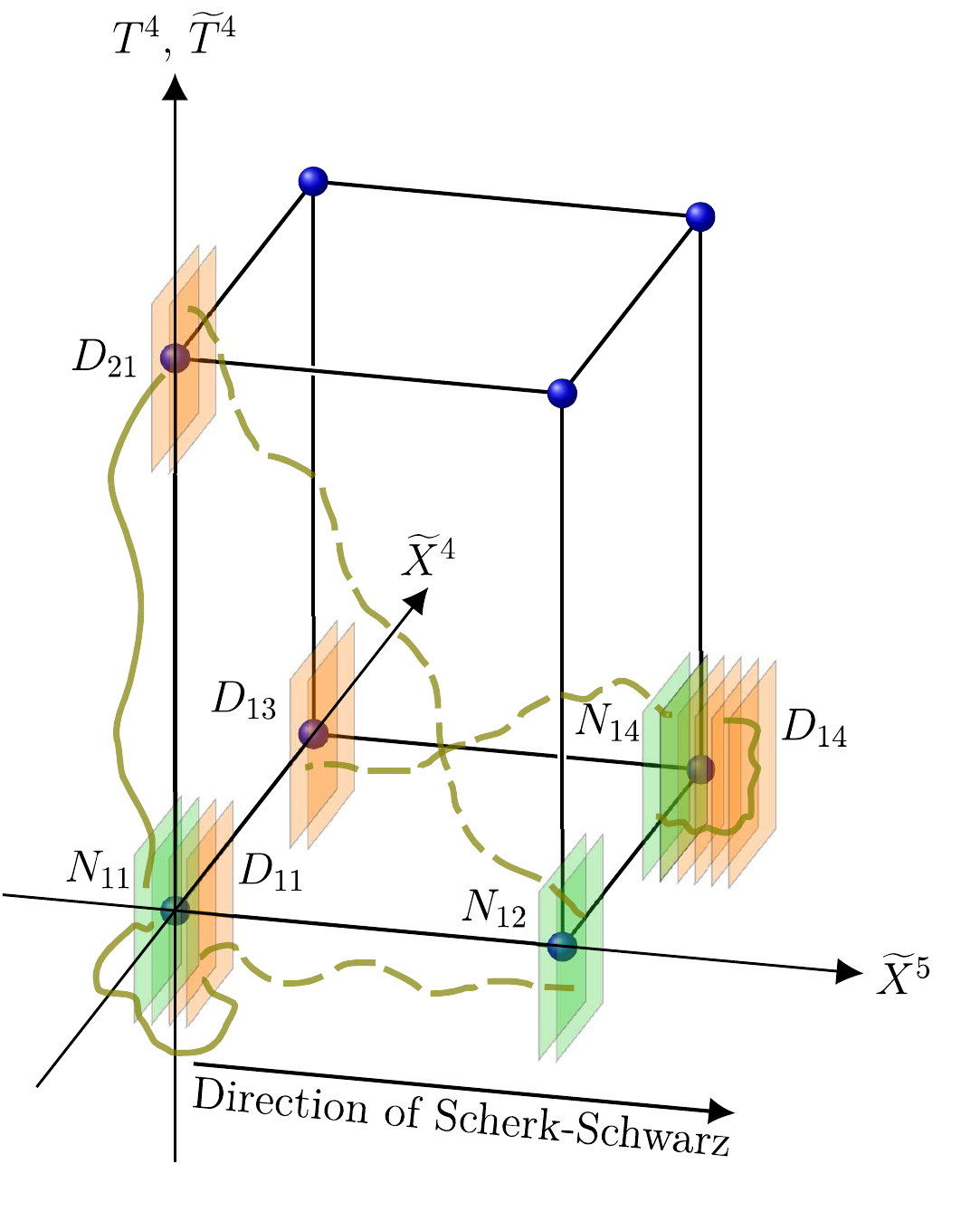}
\end{center}
\caption{ND states correspond to strings stretched between a stack of D3-branes (T-dual to D9-branes) and a stack of D3-branes (T-dual to D5-branes). They are massless bosons (solid strings) when the stacks are located on corners with common coordinates in $\tilde T^2/I_{45}$. They are fermions (dashed strings) when the corners have common coordinate $\tilde X^4$ and distinct coordinate~$\tilde X^5$. }
\label{massless_ND}
\end{subfigure}
\caption{Open string massless modes.}
\label{massless_picture}
\end{center}
\end{figure}

Massless bosons require the ends of the strings (in the D3-brane picture) to be located at fixed points $ii'$ and $jj'$ satisfying $i'=j'$. On the other hand, to be massless, the fermions need  $\vec{a}'_S+\vec{a}_{i'}-\vec{a}_{j'}=\vec{0}$ or $2\vec{a}'_S$. This is the case if the corners $ii'$ and $jj'$ are on opposite sides along the Scherk-Schwarz direction, \ie satisfying $i'=2i''-1$ and $j'=2i''$ for $i''\in \{1,2\}$, or the contrary. Moreover, for bosons and fermions  in the NN  and DD sectors to be massless, the ends of the strings (in the D3-brane picture) are  further imposed to lie at the same $\tilde T^4/\Z_2$ or $T^4/\Z_2$ position \ie $i=j$. For the states in the Neuman-Dirichlet (ND) sector however, $i$ and $j$ can be arbitrary.
To illustrate these considerations, Fig.~\ref{massless_NNDD} displays massless states arising in the NN sector (green) and DD sector (orange) that are bosonic (solid strings) or fermionic (dashed strings). Similarly, Fig.~\ref{massless_ND} shows massless strings in the ND sector (khaki) which are bosonic (solid strings) or fermionic (dashed strings).  

One can perform a precise counting of the representations under each  unitary gauge group factor. For the bosons, we have the bosonic content of $\N=2$ vector multiplets in the adjoint representations of the $U(n_{ii'})$ and $U(d_{ii'})$ gauge groups, and scalars of $\N=2$ hypermultiplets living in the antisymmetric $\oplus$ 
$\overline{\text{antisymmetric}}$ representations of $U(n_{ii'})$ and $U(d_{ii'})$. We also have scalars of hypermultiplets in the ND sector, which are in bifundamental representations of $U(n_{ii'})\times U(d_{ji'})$. The massless fermions in the  NN , DD   and ND sectors are those of hypermultiplets, all in various bifundamental representations of unitary gauge groups supported on stacks of D3-branes separated along the T-dual Scherk-Schwarz direction (and possibly along $T^4/\Z_2$ or $\tilde T^4/\Z_2$ for the ND states).

In the closed string sector, all fermions initially massless in the BSGP model acquire a mass $M$ after implementation of the Scherk-Schwarz mechanism.  The massless spectrum thus reduces to the bosonic one encountered in the BSGP model. We  obtain a total of \be
\nB^{\text{closed}}=4\times 24~,\qquad \nF^{\text{closed}}=0
\ee
fermionc and bosonic degrees of freedom. Taking into account both the closed string and open string sectors, the numbers $\nF$ and $\nB$ of massless fermionic and bosonic degrees of freedom in the $\N=2\to \N=0$ model that includes discrete WL deformations satisfy
\begin{align}
\begin{split}
\label{nfnb}
\nF-\nB=4\Big[8&-2\sum_{i,i''}\left(n_{i,2i''-1}-n_{i,2i''}\right)^{2}-2\sum_{i,i''}\left(d_{i,2i''-1}-d_{i,2i''}\right)^{2}\\
&-\sum_{i,i'',j}\left(n_{i,2i''-1}-n_{i,2i''}\right)\left(d_{i,2i''-1}-d_{j,2i''}\right)\Big]~.
\end{split}
\end{align}

%%%%%%%%%%%%%%%%%%%%%%%%%%%%%%%%%%%%%%

\section{Effective potential}
\label{sc}

In this section, we consider the model described in the previous section at points in moduli space corresponding to discrete values of the WL's. In these backgrounds, the one-loop Colemann-Weinberg effective potential is extremal with respect to the WL's, and the quantum mass terms of these moduli can be determined by Taylor expansion. However, in order to determine the true mass matrix, this computation must be supplemented by the analysis of a generalized Green-Schwarz mechanism \cite{GimonPolchinski2}, which implies anomalous $U(1)$ gauge bosons to actually be massive at tree-level. This mechanism also induces masses to twisted moduli in the closed string sector. Moreover, we will see that the internal metric (except the component involved in $M$ when $\nF\neq \nB$ ) and two-form moduli are flat directions of the one-loop potential, up to exponentially suppressed corrections. 
Finally, moduli in the ND sector may  exist. When this is the case, their masses can be determined by computing two-point functions at one loop of ``boundary changing vertex operators''. However, this is a highly non-trivial task which  will be presented in a forthcoming work~\cite{wip}.

%%%%%%

\subsection{Wilson line mass terms and effective potential}
\label{mT1}

Following the method of \cite{PreviousPaper, Paper}, the WL mass terms can be found from the one-loop Coleman-Weinberg effective potential $\V$. The latter contains contributions coming from the closed-string worldsheet topologies (torus and Klein bottle) as well as from the open string amplitudes (annulus and M\"obius strip). Because we are interested in the expansion of the potential with respect to the WL's/positions in configurations where the D3-branes are located on the orientifold planes, it is convenient to describe the WL's as fluctuations $\epsilon_{\alpha}$ and $\xi_{\alpha}$ around such backgrounds. More precisely, we define
\begin{align}
\begin{split}
\label{WLbis}
a_{\alpha}^{\I}&~=~\langle a_{\alpha}^{\I}\rangle+\epsilon_{\alpha}^{\I}~,\qquad \langle a_{\alpha}^{\I}\rangle~\in~\left\{0,\frac{1}{2}\right\},\qquad b_{\alpha}^{\I}~=~\langle b_{\alpha}^{\I}\rangle+\xi_{\alpha}^{\I}~,\qquad  \,\langle b_{\alpha}^{\I}\rangle~\in~\left\{0,\frac{1}{2}\right\}\,.
\end{split}
\end{align}
As mentioned in the introduction, we are interested in regions of moduli space where the KK mass scale associated with the large Scherk-Schwarz direction $X^5$ is lower than the string scale as well as all other mass scales induced by the compactification moduli $G_{\I\J}$. In this case, the effective potential takes the form \cite{Paper}
\begin{equation}
\V~=~\frac{\Gamma\big(\frac{5}{2}\big)}{\pi^{\frac{13}{2}}}M^{4}\sum_{l_{5}}\frac{\N_{2l_{5}+1}(\epsilon,\xi,G)}{|2l_{5}+1|^{5}}+\O\!\left((\Ms M)^{2}e^{-2\pi c\frac{\Ms }{M}}\right)\,,
\label{pofi}
\end{equation}
where $c$ is a positive constant of order 1. The quantity $\N_{2l_{5}+1}(\epsilon,\xi,G)$ captures the contribution of the potential coming from the lightest states, which correspond to KK modes propagating along $X^5$. The other states being supermassive compared to the supersymmetry breaking scale $M$, they yield exponentially suppressed contributions.

In order to find the mass terms of the $\epsilon^{\I}$ and $\xi^{\I}$, one must expand $\N_{2l_{5}+1}(\epsilon,\xi,G)$ up to quadratic order, and restrict the result to the dynamical WL degrees of freedom. As previously said, for the D3-branes T-dual to the D5-branes, there are $16$ independent positions along $\tilde T^2/I_{45}$, and at most $8$ positions in $T^{4}/\Z_2$. A similar counting is  valid for the D3-branes T-dual to the D9-branes. We label the dynamical positions in $\tilde T^{2}/I_{45}$ with an index $r'$, and in $T^{4}/\Z_2$ or $\tilde T^4/\Z_2$ with an index $r$,  
\begin{equation}
\label{epxi}
\begin{alignedat}{3}
&\epsilon_{r}^{I}~,\quad && I=6,\dots,9,\quad && r=1,\dots,\sum_{i,i'}\left\lfloor\frac{N_{ii'}}{4}\right\rfloor=\sum_{i,i'}\left\lfloor\frac{n_{ii'}}{2}\right\rfloor\leq 8-{\tilde \cR\over 2}~,\\
&\xi_{r}^{I}~,\quad && I=6,\dots,9,\quad && r=1,\dots,\sum_{i,i'}\left\lfloor\frac{D_{ii'}}{4}\right\rfloor=\sum_{i,i'}\left\lfloor\frac{d_{ii'}}{2}\right\rfloor\leq 8-{\cR\over 2}~,\\
&\epsilon_{r'}^{I'}~, ~\xi_{r'}^{I'}~,\quad && I'=4,5,\quad &&r'=1,\dots,16~.
\end{alignedat}
\end{equation}
It is convenient to denote respectively by $i(r)i'(r)$ and $j(r)j'(r)$ the corners in the appropriate T-dual pictures around which $2\pi \epsilon_r^I$ and $2\pi \xi_r^I$ fluctuate,  and by $i(r)\hat{\imath}'(r)$ and $j(r)\hat{\jmath}'(r)$  the corners which are on the opposite sides of the fifth direction. Similarly, we denote respectively by 
$i(r')i'(r')$ and $j(r')j'(r')$ the corners around which $2\pi \epsilon_{r'}^{I'}$ and $2\pi \xi_{r'}^{I'}$ fluctuate, and by $i(r')\hat{\imath}'(r')$ and $j(r')\hat{\jmath}'(r')$ the corners which are on the opposite sides of the fifth direction. In these notations, the result reads
\begin{align}
%\begin{split}
\N_{2l_{5}+1}(\epsilon,\xi,G)&\;= ~\nF-\nB+32\pi^{2}(2l_{5}+1)^{2}\Bigg\{ \nonumber \\
&\;~~\sum_{r}\Big(n_{i(r)i'(r)}-n_{i(r)\hat \imath'(r)}-1\Big)\epsilon_{r}^{I}\Delta^{IJ}\epsilon_{r}^{J}+\sum_{r}\Big(d_{j(r)j'(r)}-d_{j(r)\hat \jmath'(r)}-1\Big)\xi_{r}^{I}\Delta_{IJ}\xi_{r}^{J}\nonumber \\
&+\sum_{r'}\bigg(n_{i(r')i'(r')}-n_{i(r')\hat \imath'(r')}-1+\frac{1}{4}\sum_{i}\left(d_{ii'(r')}-d_{i\hat \imath'(r')}\right)\bigg)\epsilon_{r'}^{I'}\Delta^{I'J'}\epsilon_{r'}^{J'}\label{N_final} \\
&+\sum_{r'}\bigg(d_{j(r')j'(r')}-d_{j(r')\hat \jmath'(r')}-1+\frac{1}{4}\sum_{j}\left(n_{jj'(r')}-n_{j\hat \jmath'(r')}\right)\bigg)\xi_{r'}^{I'}\Delta^{I'J'}\xi_{r'}^{J'}\nonumber\\
&+\O\!\left(\epsilon^{4},\xi^{4}\right)\Bigg\}\;,\nonumber 
%\end{split}
\end{align}
where the delta tensors involve the metric $G$, and can be found in Ref.~\cite{Paper}. Because these tensors have positive eigenvalues, the signs of the mass terms are those of the pre-factors, in parentheses. Note that, with techniques similar to those described in \cite{SNS1, SNS2, CoudarchetPartouche,PreviousPaper,Paper}, these pre-factors can be determined by simple algebraic computations using the sole knowledge of the massless spectrum and their  representations. They can be expressed  in terms of Dynkin indices. Notice that Eq.~(\ref{N_final}) shows explicitly that the backgrounds under considerations are extrema of the potential.

Inspecting Eq.~(\ref{N_final}), one finds that the mass terms of $\epsilon_r^I$ and $\xi_r^I$ are non-negative if and only if the brane configuration satisfies
\be
\forall~i\,,i'':~ (n_{i,2i''-1},n_{i,2i''})~,~ (d_{i,2i''-1},d_{i,2i''})\in \big\{(0,p),\,(p,0),\,(1,p),\,(p,1)~\,\where\,~p\in\mathbb{N} \big\}\,.
\label{condiwl}
\ee
However,  the mass terms of $\epsilon_{r'}^{I'}$ and $\xi_{r'}^{I'}$ are not enough to conclude in general on the stability/instability of these moduli. Indeed, as will be seen in the next subsection, some combinations of these scalars acquire a tree-level mass thanks to a generalized Green-Schwarz mechanism.

\subsection{Mass generation {\bf \em via} generalized Green-Schwarz mechanism}
\label{GSm} 

Since all $\N=1$ supersymmetric theories in six dimensions are chiral, anomaly cancellations in the BSGP type IIB orientifold model on $T^4/\Z_2$ proceed in a non-trivial way. For any values of the WL's along  $T^4/\Z_2$  for the D9-brane gauge group, and arbitrary positions of  the D5-branes in $T^4/\Z_2$, the fermionic spectrum ensures the cancellation of the irreducible gauge and gravitational anomalies. However, there are residual reducible anomalies, which are described by an anomaly polynomial $I_8$ explicitly written down  in~\cite{GimonPolchinski2}. When the WL's and positions take discrete values $\vec a_i$, the gauge symmetry generated by the D9-branes and D5-branes is a product of unitary groups,
\be
\prod_{i/n_i\neq 0} U(n_i)\times \prod_{j/d_j\neq 0}U(d_j)~,\quad \where \quad \sum_i n_i=\sum_i d_i=16~,
\label{uni}
\ee 
and where the rank is 32.  As usual in six dimensions, the anomaly polynomial $I_8$ does not factorise, reflecting the fact that massless forms transform nonlinearly under  gauge transformations and diffeomorphisms. In the case at hand, these forms are RR fields belonging to the closed string spectrum: there is the two-form $C$ in the untwisted sector, as well as sixteen four-forms $C^i_4$ in the twisted sector. By Hodge duality ($\dd C^i_4=*\dd C^i_0$), the magnetic four-form degrees of freedom are equivalent to electric pseudoscalars $C^i_0$. Each of them combines with 3 NS-NS scalars of the twisted sector, thus realizing the bosonic part of the massless twisted hypermultiplet localized at the fixed point $i$ of $T^4/\Z_2$.

Anomaly cancellation requires the effective action to contain tree-level couplings proportional to 
\be
\int C \wedge X_4~~  \quad \mbox{or} ~~\quad \sum_{i,a}c_{ia}  \int C_0^i \wedge F^3_a + \sum_{i,a}c_{ia}  \int C_4^i \wedge F_a ~ , \label{mass2}
\ee
where $F_a$, $a=1,\dots,16$, are the field strengths of the Cartan $U(1)$ generators of $\prod_{i/d_i\neq 0} U(d_i)$, while $F_a$, $a=17,\dots,32$, are those of $\prod_{i/n_i\neq 0} U(n_i)$. Similar couplings involving ${\rm tr} \,R^2$ also exist.  In the above expressions, the coefficients are~\cite{GimonPolchinski2, Paper} 
\be
\begin{aligned}
c_{ia}&=4\delta_{a\in i}~,&&\mbox{for $a=1,\dots,16$}~,\\
c_{ia}&=-e^{4i\pi \vec a_i\cdot \vec a_{j(a)}}~, &&\mbox{for $a=17,\dots,32$}~,
\end{aligned}
\ee
where $\delta_{a\in i}=1$ when the  $a$-th $U(1)$ belongs to the Cartan subalgebra of $U(d_i)$, and $\delta_{a\in i}=0$ otherwise. Moreover, we denote by $2\pi \vec a_{j(a)}$ the coordinate vector of the corner of $\tilde T^4/\Z_2$ which supports the Cartan $U(1)$  labelled by $a$ of  $\prod_{j/n_j\neq 0} U(n_j)$  (in a T-dual description).
The Lagrangian can be cast into a local form by dualizing the last term in Eq.~(\ref{mass2}),
which becomes
\be
\sum_{i}\int \big(C_0^i +\sum_{a} c_{ia} A_a\big) \wedge * \big(C_0^i + \sum_{b}c_{ib} A_b\big)~, 
\label{mass3}
\ee
where the $A_a$ denote the Abelian vector potentials, $F_a=\dd A_a$. As a result, the latter admit a tree-level mass term 
\be
\half \sum_{a,b}A_a \M_{ab}^2 A_b~,\quad \where \quad \M_{ab}^2=\sum_i c_{ia}c_{ib}~.
\ee
The mass matrix $\M^2$ can be diagonalized by an orthogonal transformation, $A_a = \P_{ab} \hat A_b$. Denoting the eigenvalues by $\M^2_a$, the nonzero ones (which are actually positive) are in one-to-one correspondence with  the Stueckelberg fields $C^i_0$ which are eaten by the $\hat A_a$'s that gain a mass. One can see that if there are 16 or fewer unitary factors in Eq.~(\ref{uni}), all of them are broken to $SU$ groups, while if there are more than 16 unitary factors, exactly 16 are broken to $SU$ groups~\cite{GimonPolchinski2}. By supersymmetry, all twisted hypermultiplets initially containing the $C^i_0$'s which are eaten also become massive. They combine with Abelian vector multiplets to become long massive vector multiplets. As a result, there are between 2 and 16 twisted quaternionic scalars for which stability is automatically guaranteed.   

Compactifying down to four dimensions, we may define the WL's along $T^2$ as $\hat A^{I'}_a=\hat \xi^{I'}_a$,  and write their total mass terms by adding the tree-level contributions to the one-loop effective potential corrections, 
\be
\hat \xi^{I'}_d\left[\M^2_d\,\delta_{dc}\,\delta_{I'J'}+ \P_{ad}\,{\partial V\over \partial \xi^{I'}_a\partial \xi^{J'}_b}\,\P_{bc}\right] \hat \xi^{J'}_c~,
\ee
where $(\xi^{I'}_1,\dots,  \xi^{I'}_{32})\equiv (\xi^{I'}_1,\dots,  \xi^{I'}_{16},\epsilon^{I'}_1,\dots,  \epsilon^{I'}_{16})$. In the above formula, both contributions are proportional to the open string coupling. However, while the first one is a supersymmetric mass term proportional to $\Ms^2$, the second one scales like $(M^2/\Ms)^2$, which is always subdominant in the regime $M<\Ms$. Hence, all  WL's of massive $\hat A_a$'s are super heavy and can be safely set to zero in a study of moduli stability,
\be
\xi^{I'}_a\equiv 0~,\quad \mbox{when $\M^2_a>0$}~.
\ee
For the remaining WL's denoted $\xi^{I'}_u$ to be non-tachyonic at one-loop, one needs to find brane configurations such that the mass matrix 
\be
 \P_{au}\,{\partial V\over \partial \xi^{I'}_a\partial \xi^{J'}_b}\,\P_{bv}~,\quad \mbox{for $u,v$ such that $\M^2_u,\M^2_v= 0$}~, 
 \ee
has non-negative eigenvalues. 

%%%

\subsection{Closed string moduli}

We have already mentioned that 2 to 16 of the twisted quaternionic moduli acquire a mass \via the Green-Schwarz mechanism described in the previous subsection. We have not computed the masses of the remaining twisted closed-string moduli that may be determined by evaluating two-point functions. 

In the untwisted sector, when the D3-branes sit on the O3-planes, all fluctuations in Eq.~(\ref{N_final}) vanish and the potential reduces to Eq.~(\ref{rule_of_thumb}), which does not involve the metric components $G_{\I\J}$. Hence, up to exponentially suppressed corrections, all components $G_{\I\J}$ are flat directions, except $G^{55}$ (which appears in the definition of $M$) when $\nF-\nB\neq 0$.  Moreover, the RR two-form $C_{\I\J}$ is mapped by heterotic-type I duality to the antisymmetric tensor $B_{\I\J}$. In four dimensions, the duality is at weak coupling on both sides~\cite{dual0,dual1,dual2,dual3}. In the heterotic theory, when $M$ is lower than all  other mass scales, the dependence of the Colemann-Weinberg effective potential on $B_{\I\J}$ arises from loops of generically massive states that become massless at special points in moduli space. The key point is that these states have non-trivial winding numbers along the internal directions and are therefore mapped to D1-branes on the type~I side. As a result, up to exponentially suppressed corrections, the one-loop effective potential does not depend on $C_{\I\J}$, implying these moduli to be   flat directions.

%%%%%%%%%%%%%%%%%%%%%%%%%%%%%%%%%%%%%

\section{Stability analysis of the models}
\label{models}

Let us now analyze  the stability of the backgrounds at one loop. As said earlier, we do not compute in the present work the quantum masses of the moduli in the ND sector. However,  the absence of such fields is ensured when the D3-branes associated with the D9-branes and those associated with the D5-branes never share the same position in $\tilde{T}^2/I_{45}$
\begin{equation}
\mbox{\em no moduli in the ND sector:}\quad n_{ii'}d_{ji'}=0~\quad\forall i,j,i'~.
\end{equation}
In the sequel, we first explore in detail  models belonging to the non-pertubatively consistent components of the moduli space $(\cR,\tilde \cR)=(0,0)$ and $(\cR,\tilde \cR)=(16,16)$ to familiarize with the implementation of the Green-Schwarz mechanism. Then, thanks to a numerical exploration of all possible brane configurations, we  list all setups that yield vanishing or positive one-loop potentials that are tachyon free  (up to exponentially suppressed terms).

%%%

\subsection{\bm Component $(\cR,\tilde \cR)=(0,0)$} 

In this component of the moduli space, there is no brane with rigid position in $T^4/\Z_2$ or $\tilde T^4/\Z_2$. Let us analyze the simplest configuration where all D3-branes T-dual to the D5-branes are located at the same fixed point $i_0$ of $T^4/\Z_2$, and similarly the D3-branes T-dual to the D9-branes are coincident on the fixed point $j_0$ of $\tilde T^4/\Z_2$. In six dimensions, the open string gauge group is $U(16)\times U(16)$. 

In order to write the classical mass matrix squared $\M^2$ of the vector potentials $A_{a},\ a=1,\dots,32$, of the Cartan subgroup (see  Sect.~\ref{GSm}), it is convenient to label by $r'\equiv a=1,\dots,16$ those associated with the D5-brane gauge group, and by $\tilde r'\equiv a=17,\dots,32$ those corresponding to the D9-branes.  In these notations, the mass matrix squared reads
\begin{equation}
\label{param_M}
\M^2=\left(\begin{array}{@{}c|c@{}}
\M^2_{r's'} & \M^2_{r'\tilde{s}'} \\
\hline
\M^2_{\tilde{r}'s'} & \M^2_{\tilde{r}'\tilde{s}'}
\end{array}\right)~,
\end{equation}
where the four $16\times 16$ blocks are 
\begin{equation}
\begin{aligned}
&\M^2_{r's'}=16~,\qquad&&\M^2_{r'\tilde{s}'}=-4e^{4i\pi\vec{a}_{i_0}\cdot\vec{a}_{j_0}}~,\\
&\M^2_{\tilde{r}'s'}=-4e^{4i\pi\vec{a}_{i_0}\cdot\vec{a}_{j_0}}~,\qquad&&\M^2_{\tilde{r}'\tilde{s}'}=16~.
\end{aligned}
\end{equation}
This matrix has two positive eigenvalues while the others vanish. As expected, two (anomalous) combinations of  Abelian vector potentials are massive, leading to  the actual $SU(16)\times SU(16)$ gauge group. The WL's along $\tilde T^2$ of the massive gauge bosons   must  be set to zero, which yields 
\begin{equation}
\xi_1^{I'}=-\sum_{r'\neq 1}\xi_{r'}^{I'}\quad\text{and}\quad\epsilon_1^{I'}=-\sum_{\tilde{r}'\neq 1}\epsilon_{\tilde{r}'}^{I'}~.
\label{mass}
\end{equation}

\begin{figure}[h!]
\captionsetup[subfigure]{position=t}
\begin{center}
\begin{subfigure}[t]{0.31\textwidth}
\begin{center}
\includegraphics [scale=0.54]{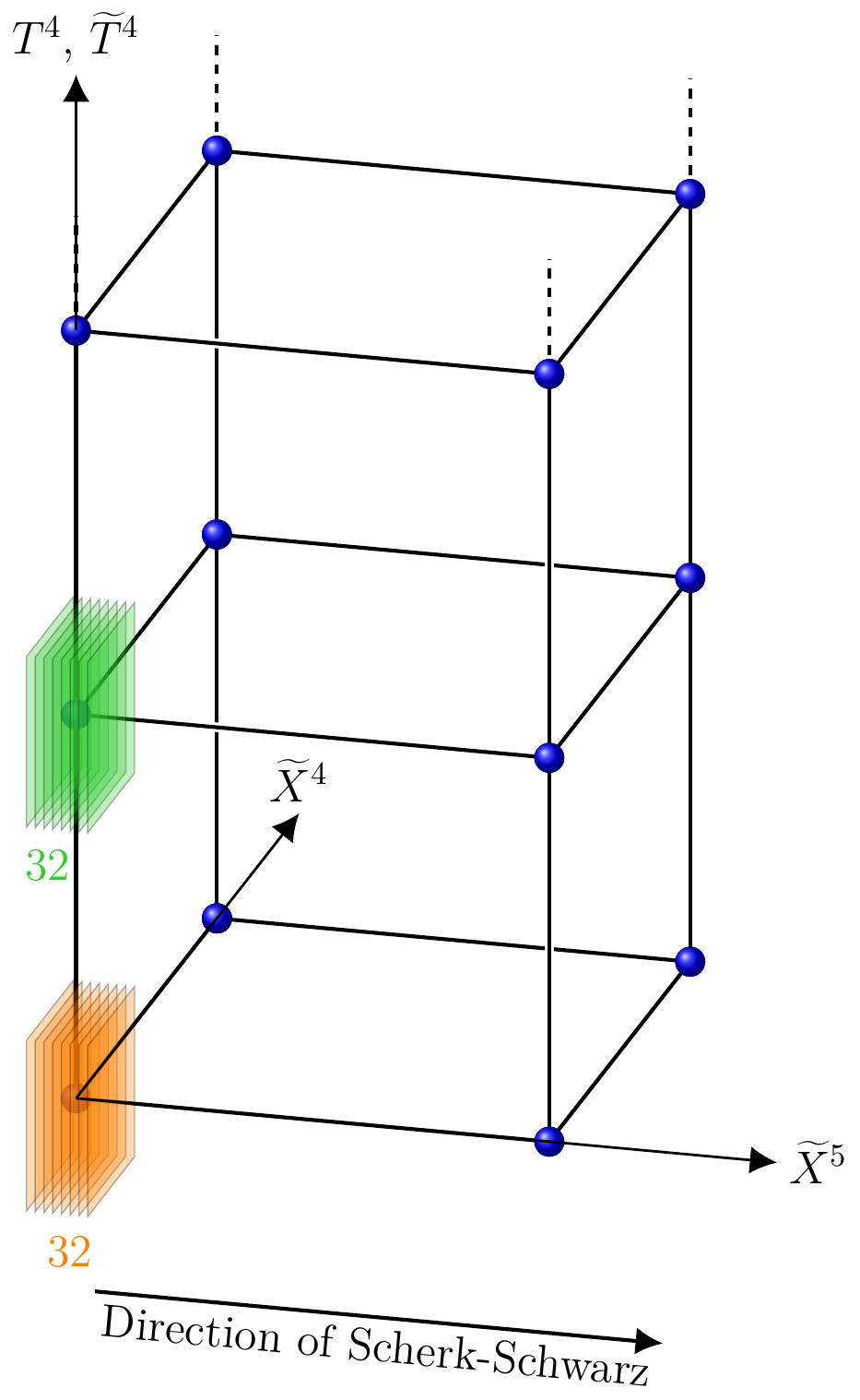}
\end{center}
\caption{\footnotesize The $32$ D3-branes associated with the D5-branes and the $32$ ones associated with the D9-branes are located at the same $\tilde T^2/I_{45}$ position.}
\end{subfigure}
\quad
\begin{subfigure}[t]{0.31\textwidth}
\begin{center}
\includegraphics [scale=0.54]{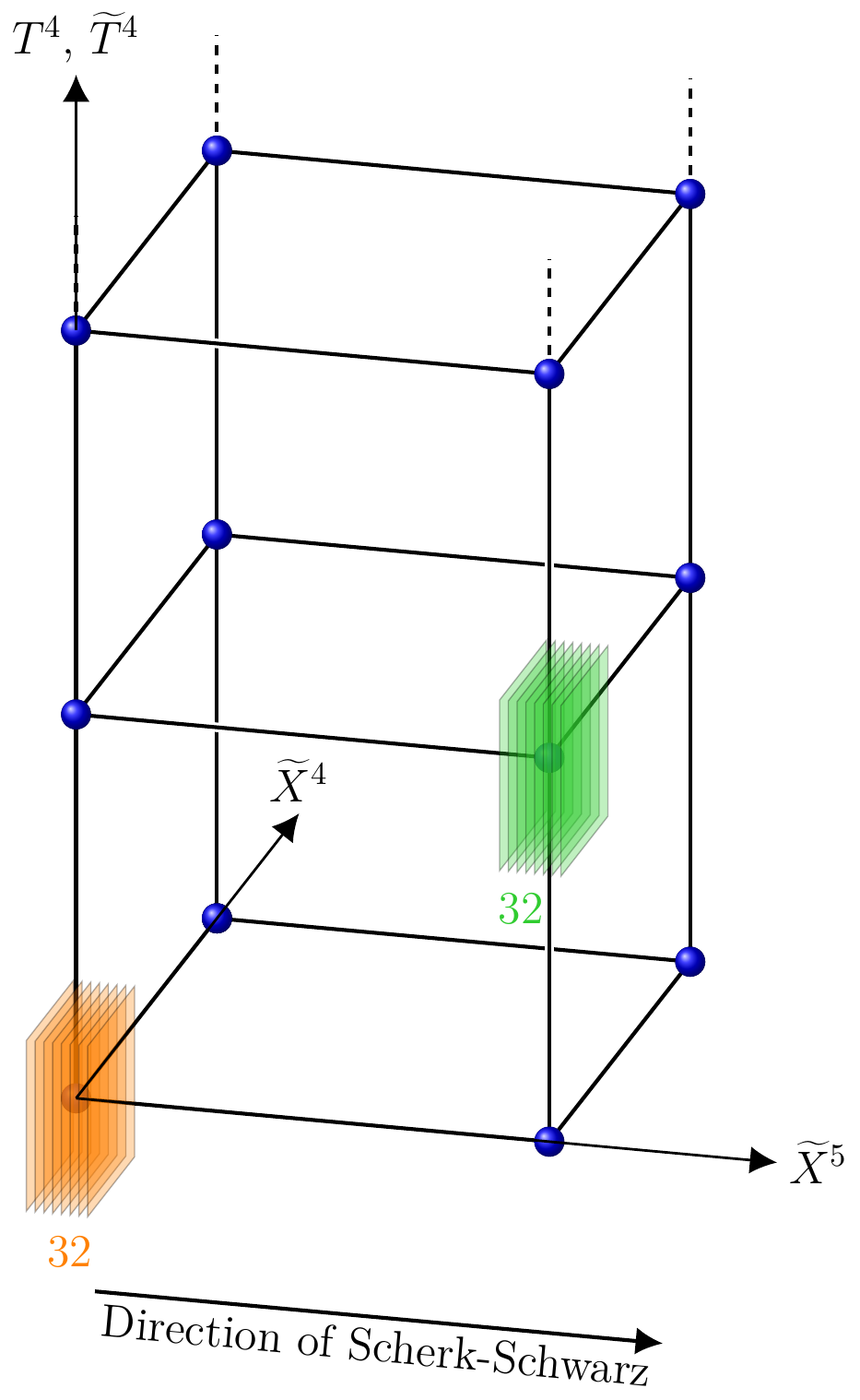}
\end{center}
\caption{\footnotesize The two stacks are located on opposite sides of the Scherk-Schwarz direction but have the same coordinate $\tilde X^4$.}
\end{subfigure}
\quad
\begin{subfigure}[t]{0.31\textwidth}
\begin{center}
\includegraphics [scale=0.54]{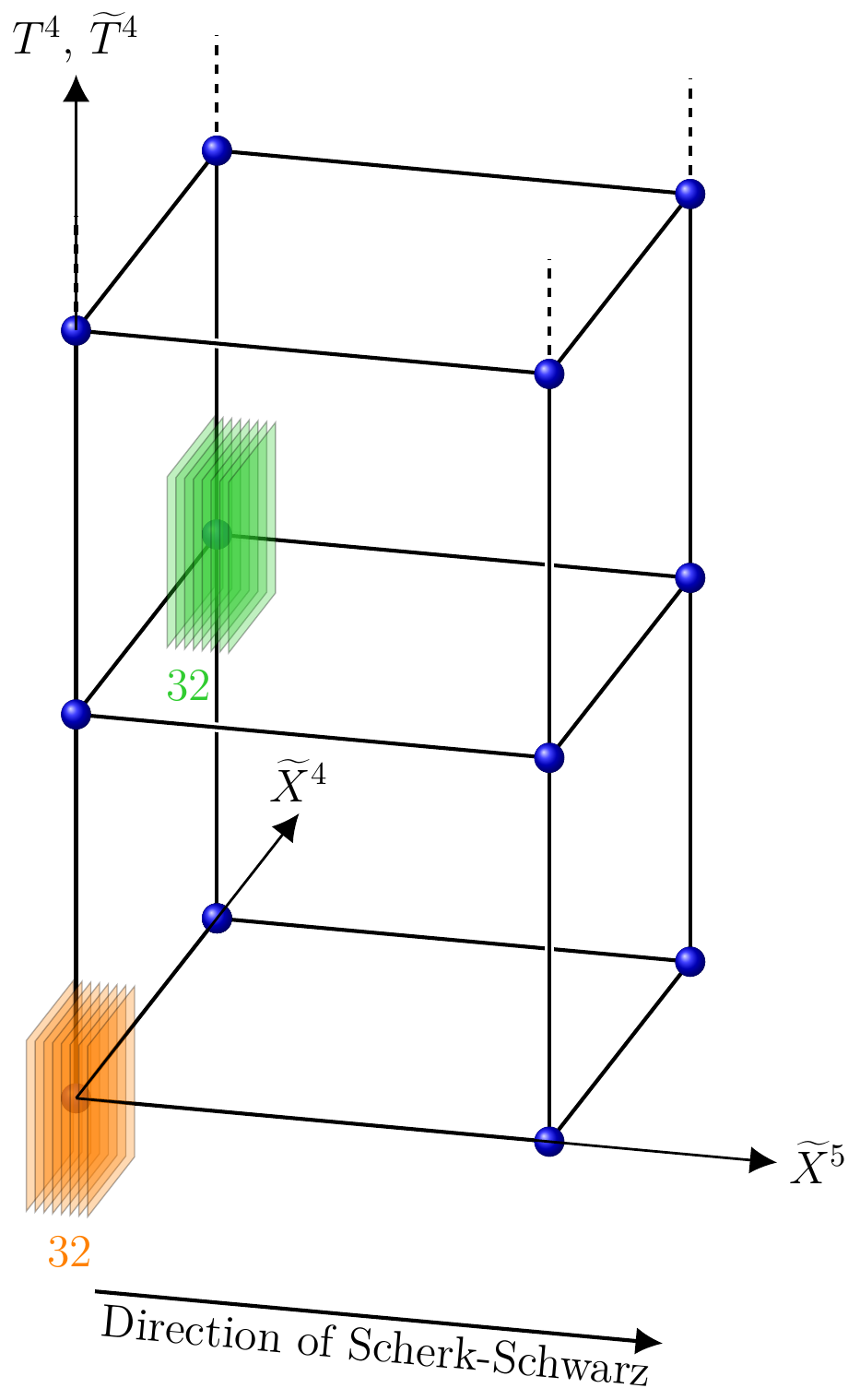}
\end{center}
\caption{\footnotesize The two stacks have distinct  coordinate $\tilde X^4$. Their positions along the Scherk-Schwarz direction is irrelevant.}
\end{subfigure}
\caption{\footnotesize Geometric representation of three brane configurations of the component $(\cR,\tilde \cR)=(0,0)$. The D3-branes T-dual to the  D5-branes (D9-branes) are located on the same fixed point of $T^4/\Z_2$ ($\tilde T^4/\Z_2$).}
\label{examples_1}
\end{center}
\end{figure}

Let us analyze in detail the case where the D3-branes associated with  the D5-branes are all located at the corner $i'_0$ of $\tilde{T}^2/I_{45}$, and similarly those corresponding to the D9-branes are coincident at the fixed point $j'_0$. The positions $\epsilon_r^I$ and $\xi_r^I$ along $\tilde T^4/\Z_2$ or $T^4/\Z_2$ all have the same positive mass term coefficient\footnote{Those integer coefficients that appear in parentheses in Eq.~(\ref{N_final}).} equal to 15  in Eq.~(\ref{N_final}), and are therefore stabilized. 
Before the Green-Schwarz mechanism is taken into account, the mass terms in Eq.~(\ref{N_final}) of the positions $\epsilon_{r}^{I'}$ and $\xi_r^{I'}$ along $\tilde T^2/I_{45}$ depend on the precise distribution of the stacks.  The mass coefficients are $(16-0-1+{\delta\over 4} \,16)=15+4\delta$, where
\begin{enumerate}[label=(\alph*)]
\item $\delta=+\ 1$ if the two stacks of branes are at the same $\tilde T^2/I_{45}$ position: $i'_0=j'_0$,
\item $\delta=-\ 1$ if the two stacks of branes have the same coordinate $\tilde X^4$ but sit on opposite sides of the Scherk-Schwarz direction: $i'_0=2i''_0-1$ and $j'_0=2i''_0$ or the contrary,
\item $\delta=0$ if the stacks do not have the same coordinate $\tilde X^4$.
\end{enumerate}
Fig.~\ref{examples_1} depicts these three possibilities.
In all cases, the WL's are therefore stabilized. However, to find the correct masses once the Green-Schwarz mechanism is taken into account, $\xi_1^{I'}$ and $\epsilon_1^{I'}$ can be eliminated thanks to  Eq.~(\ref{mass}), which yields a new mass matrix squared for the 30 remaining degrees of freedom. 

As a consequence of the Green-Schwarz mechanism, two twisted quaternionic scalars acquire a mass, while 14 remains to be dealt with. Moreover, moduli in the ND sector coming from the hypermultiplet in the bifundamental of $SU(16)\times SU(16)$ are present in Case (a), and their masses should also be analyzed in detail~\cite{wip}. In Cases (a), (b), (c), the massless spectrum yields $\nF-\nB=-4064-256\delta$ leading to a negative potential.

%%%

\subsection{\bm Component $(\cR,\tilde{\cR})=(16,16)$} 

All D3-branes have rigid positions in $T^4/\Z_2$ or $\tilde T^4/\Z_2$. They are grouped by pairs located at each fixed point of $T^4/\Z_2$ or $\tilde T^4/\Z_2$, which yields the gauge symmetry $U(1)^{16}\times U(1)^{16}$. The mass matrix squared $\M^2$ of the Abelian vector potentials, which  is given by
\begin{equation}
\begin{aligned}
&\M^2_{r's'}=16\delta_{r's'}~,\qquad&&\M^2_{r'\tilde{s}'}=-4e^{4i\pi\vec{a}_{i(r')}\cdot\vec{a}_{i(\tilde{s}')}}~,\\
&\M^2_{\tilde{r}'s'}=-4e^{4i\pi\vec{a}_{i(\tilde{r}')}\cdot\vec{a}_{i(s')}}~,\qquad&&\M^2_{\tilde{r}'\tilde{s}'}=16\delta_{\tilde{r}'\tilde{s}'}~,
\end{aligned}
\end{equation}
possesses 16 positive eigenvalues and 16 vanishing ones. Setting to zero the WL's along $T^2$ of the massive combinations allows to eliminate all $\epsilon_{r'}^{I'}$ degrees of freedom,
\begin{equation}
\label{rel_GS}
4\epsilon_{\tilde{r}'}^{I'}=-\sum_{s'}e^{4i\pi\vec{a}_{i(\tilde{r}')}\cdot\vec{a}_{i(s')}}\xi_{s'}~.
\end{equation}
Moreover, the Green-Schwarz mechanism also induces large masses to the 16 twisted quaternionic moduli, ensuring the orbifold point $T^4/\Z_2$ of the $K_3$ manifold to be stabilized. Let us now consider specific brane configurations and analyze the stability of the WL's.

\begin{figure}[H]
\captionsetup[subfigure]{position=t}
\begin{center}
\begin{subfigure}[t]{0.31\textwidth}
\begin{center}
\includegraphics [scale=0.52]{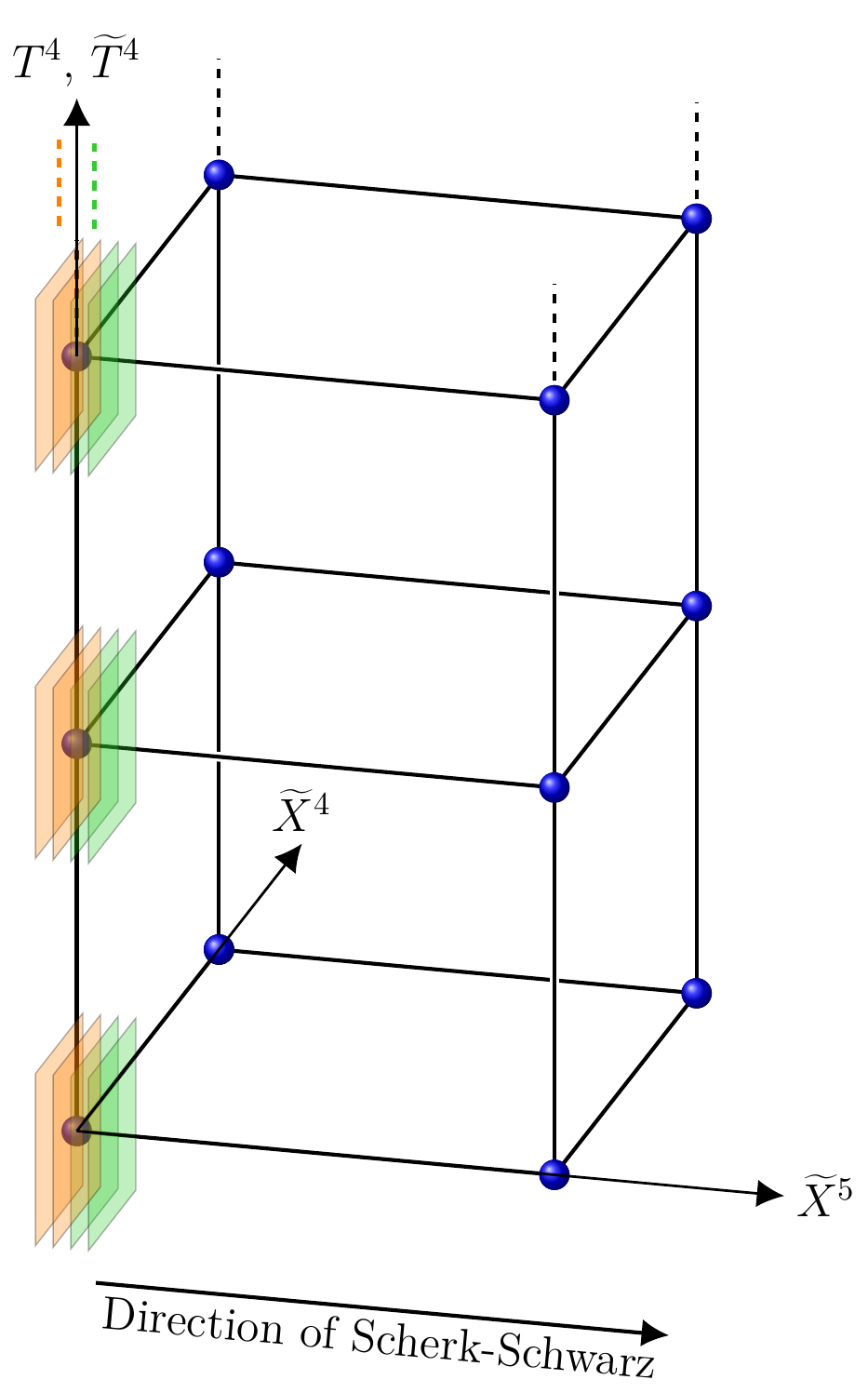}
\end{center}
\caption{\footnotesize The $16$ pairs of D3-branes associated with the D5-branes and the $16$ ones associated with the D9-branes are located at the same $\tilde T^2/I_{45}$ position.}
\label{nfnba}
\end{subfigure}
\quad
\begin{subfigure}[t]{0.31\textwidth}
\begin{center}
\includegraphics [scale=0.52]{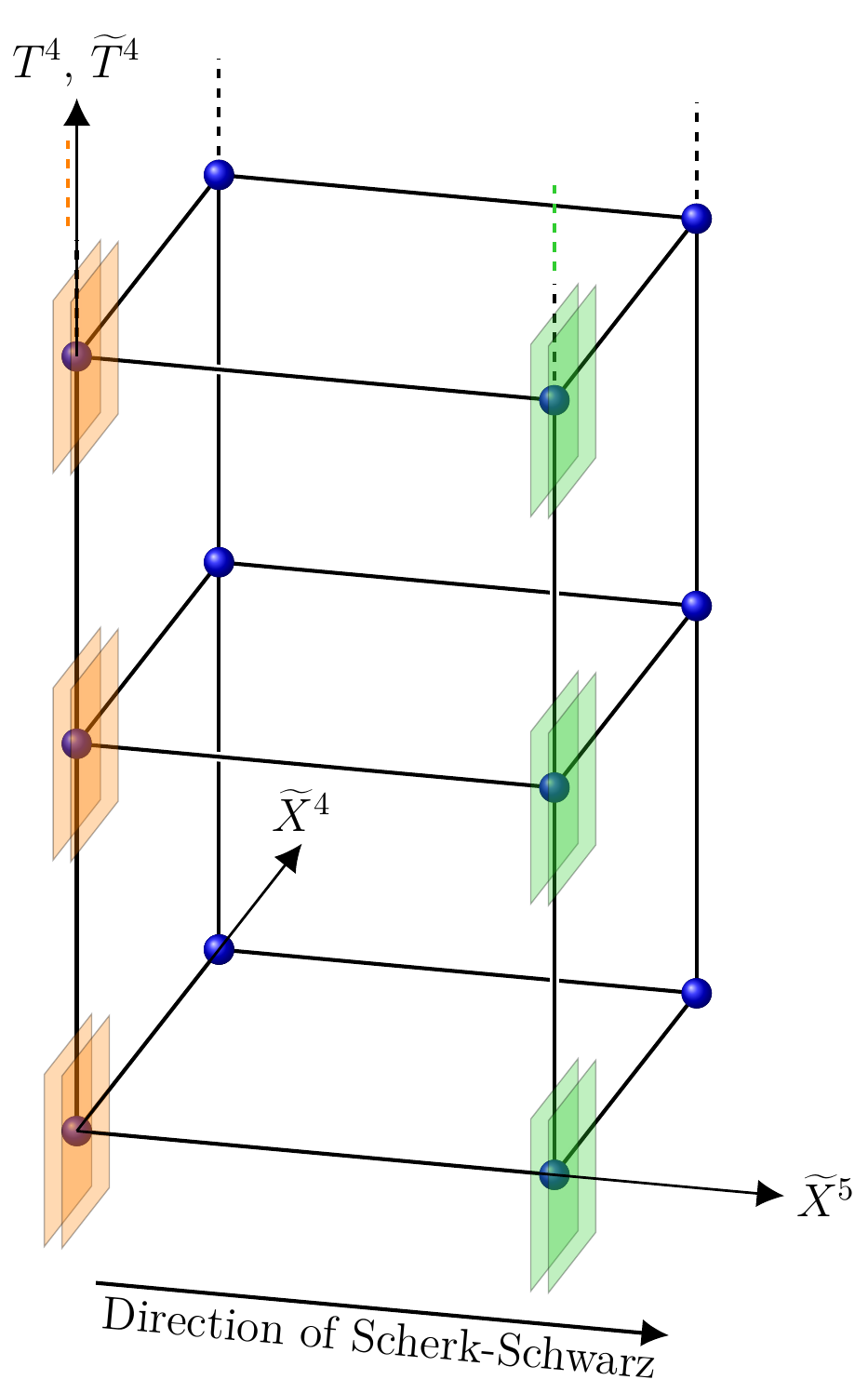}
\end{center}
\caption{\footnotesize The pairs of different kinds are located on opposite sides of the Scherk-Schwarz direction, but have the same coordinate $\tilde X^4$.}
\label{nfnbb}
\end{subfigure}
\quad
\begin{subfigure}[t]{0.31\textwidth}
\begin{center}
\includegraphics [scale=0.52]{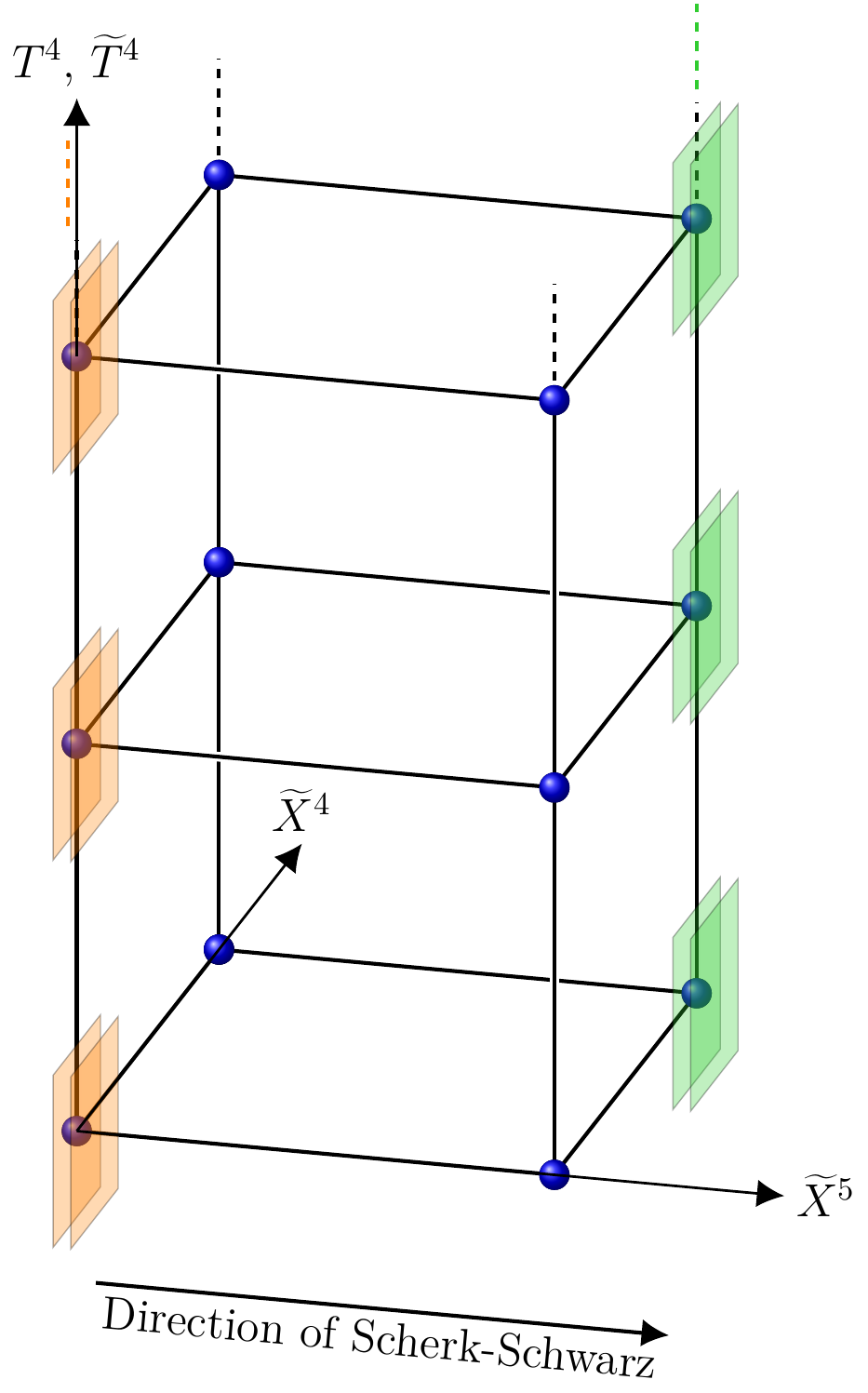}
\end{center}
\caption{\footnotesize The pairs of different kinds are located at different $\tilde X^4$ position. Their positions along the Scherk-Schwarz direction is irrelevant.}
\label{nfnbc}
\end{subfigure}
\quad
\begin{subfigure}[t]{0.31\textwidth}
\begin{center}
\includegraphics [scale=0.52]{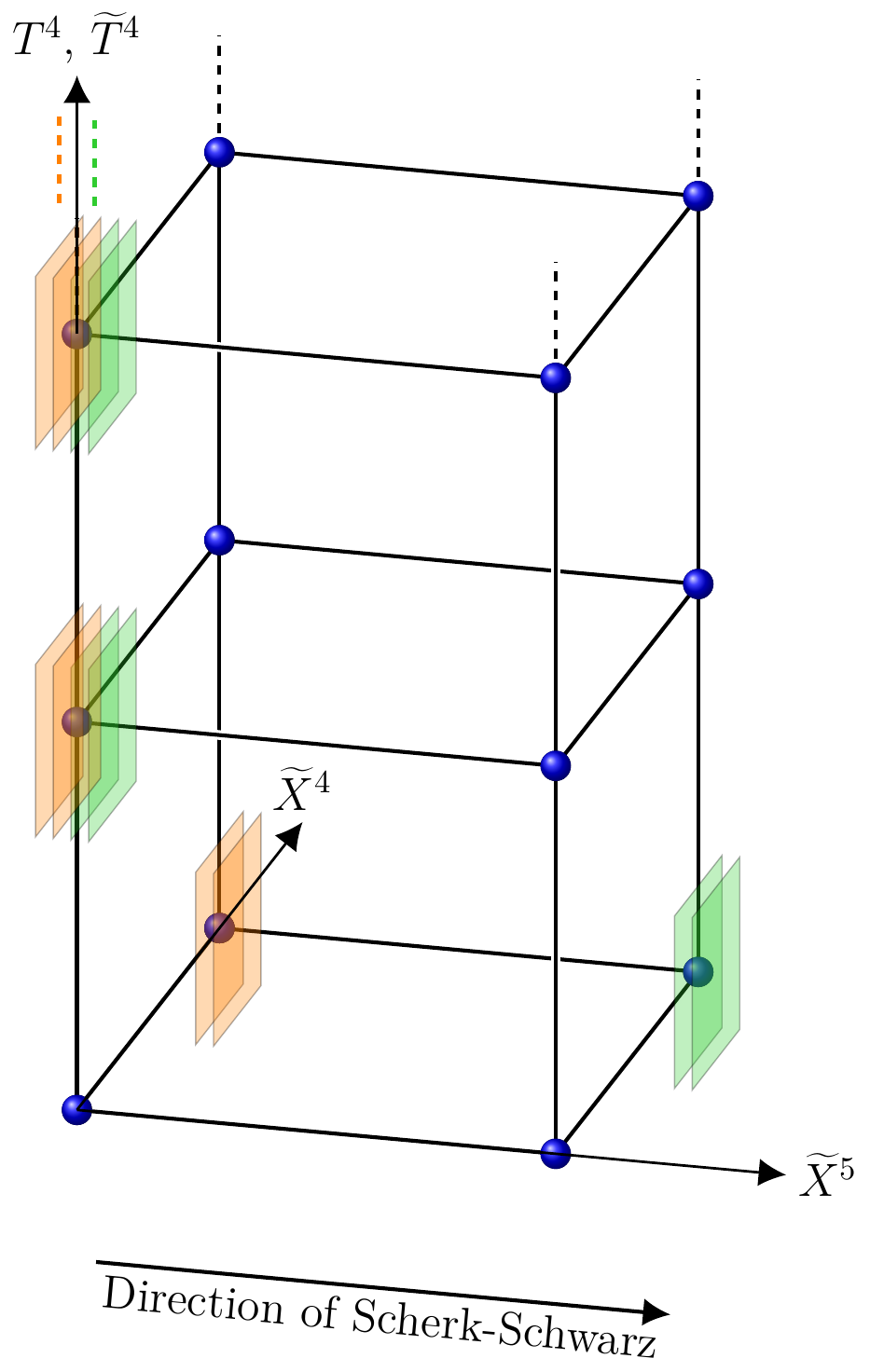}
\end{center}
\caption{\footnotesize $15$ pairs of each kind are located at the same $\tilde T^2/I_{45}$ fixed point, while the remaining pairs, displaced along $\tilde X^4$, face each other along the Scherk-Schwarz direction.}
\label{nfnbd}
\end{subfigure}
\caption{\footnotesize Geometric representations of various brane configurations in the component $(16,16)$ of the open-string moduli space.}
\label{examples_2}
\end{center}
\end{figure}

\paragraph{\em Example 1:} The simplest example amounts to puting all D3-branes T-dual to the D5-branes at a same $\tilde T^2/I_{45}$ fixed point $i'_0$,  and similarly all D3-branes T-dual to the D9-branes at the same fixed point $j'_0$, so that
\begin{equation}
n_{i,i'_0}=1,\ \forall i\in\{1,\dots,16\}~~\quad\text{and}~~\quad d_{i,j'_0}=1,\ \forall i\in\{1,\dots,16\}~.
\end{equation} 
Again, three cases (a), (b) and (c) are allowed, corresponding to having respectively the two kinds of branes at the same fixed points of $\tilde T^2/I_{45}$  ($i'_0=j'_0$), or facing each other along the Scherk-Schwarz direction ($i'_0=2i''_0-1,\ j'_0=2i''_0$ or the contrary), or finally having different coordinates along  $\tilde X^4$, as shown   in Figs~\ref{nfnba}--\ref{nfnbc}.

\begin{itemize}
\item In Case (a), all mass terms of the WL's along $T^2$ are strictly positive in Eq.~(\ref{N_final}),  implying these moduli to be stabilized. However, there are  $16^2$ massless quaternionic scalars arising from the ND sector, whose masses at one-loop  should be analyzed in order to determine whether the configuration is stable or not. The potential is negative, with $\nF-\nB=-1248$.
\item In Case (b), the mass terms of the WL's along $T^2$ are all strictly negative in Eq.~(\ref{N_final}).  Hence, the brane configuration is unstable. Before condensation of the moduli, the potential is positive, with $\nF-\nB=800$.
\item In Case (c), all mass terms of the WL's along $T^2$ vanish in Eq.~(\ref{N_final}). After elimination of the $\epsilon^{I'}_{\tilde r'}$ thanks to Eq.~(\ref{rel_GS}), all moduli $\xi^{I'}_{r'}$ remain massless. In fact, it turns out that (up to exponentially suppressed terms) the one-loop effective potential does not depend on these moduli, which  are therefore flat directions. As in Case (b), there are no moduli in the ND sector. The potential is negative, with  $\nF-\nB=-224$. 
\end{itemize}

\paragraph{\em Example 2:} More involved configurations can be considered  where the pairs of branes of each typess are distributed at different fixed points of $\tilde T^2/I_{45}$. It is then possible to find stable brane configurations even when mass terms in Eq.~(\ref{N_final}) are negative. For example, consider Case (d) in Fig.~\ref{nfnbd}. There are $15+15$ pairs of D3-branes T-dual to D5-branes  or D9-branes at a given fixed point of $\tilde T^2/I_{45}$, while the two remaining pairs are displaced along $\tilde X^4$ and face each other along $\tilde X^5$. 
The mass-term coefficients of the WL's along $T^2$ and associated with the $2\times 15$ pairs of branes are $\frac{15}{4}$, while those associated with the two remaining ones are $-\frac{1}{4}$. It turns out that eliminating the 16 $\epsilon_{r'}^{I'}$  thanks to Eq.~(\ref{rel_GS}) yields a new $16\times 16$ mass matrix with only positive eigenvalues. As a result, the brane configuration is stable, provided the $15^2$ quaternionic  moduli arising from  the ND sector do not introduce instabilities. The magnitude of the potential in this case is given by $\nF-\nB=-1120$.

%%%%%%%%%

\subsection{Full scan of the six components of the moduli space}

As explained in the introduction, configurations that are tachyon free at one loop, with positive or vanishing (up to exponentially small contributions) potentials are expected to be rare. For instance, for toroidal compactifications in dimension $d\ge 5$, it is shown in Refs~\cite{PreviousPaper,APP} that there exists only one  orientifold model (with non-exotic orientifold planes) consistent non-perturbatively, tachyon free at one loop and with non-negative potential. It is defined  in five dimensions, has a trivial open string gauge group $SO(1)^{32},$\footnote{$SO(1)$ denotes the group containing only the neutral element.}  and satisfies $\nF-\nB=8\times 8$. By a computer scan of all possible brane configurations in the orbifold case (see Ref.~\cite{Paper} for more details), we show that  few more examples exist and we describe them all.

\begin{figure}[H]
\captionsetup[subfigure]{position=t}
\begin{center}
\begin{subfigure}[b]{0.31\textwidth}
\begin{center}
\includegraphics [scale=0.54]{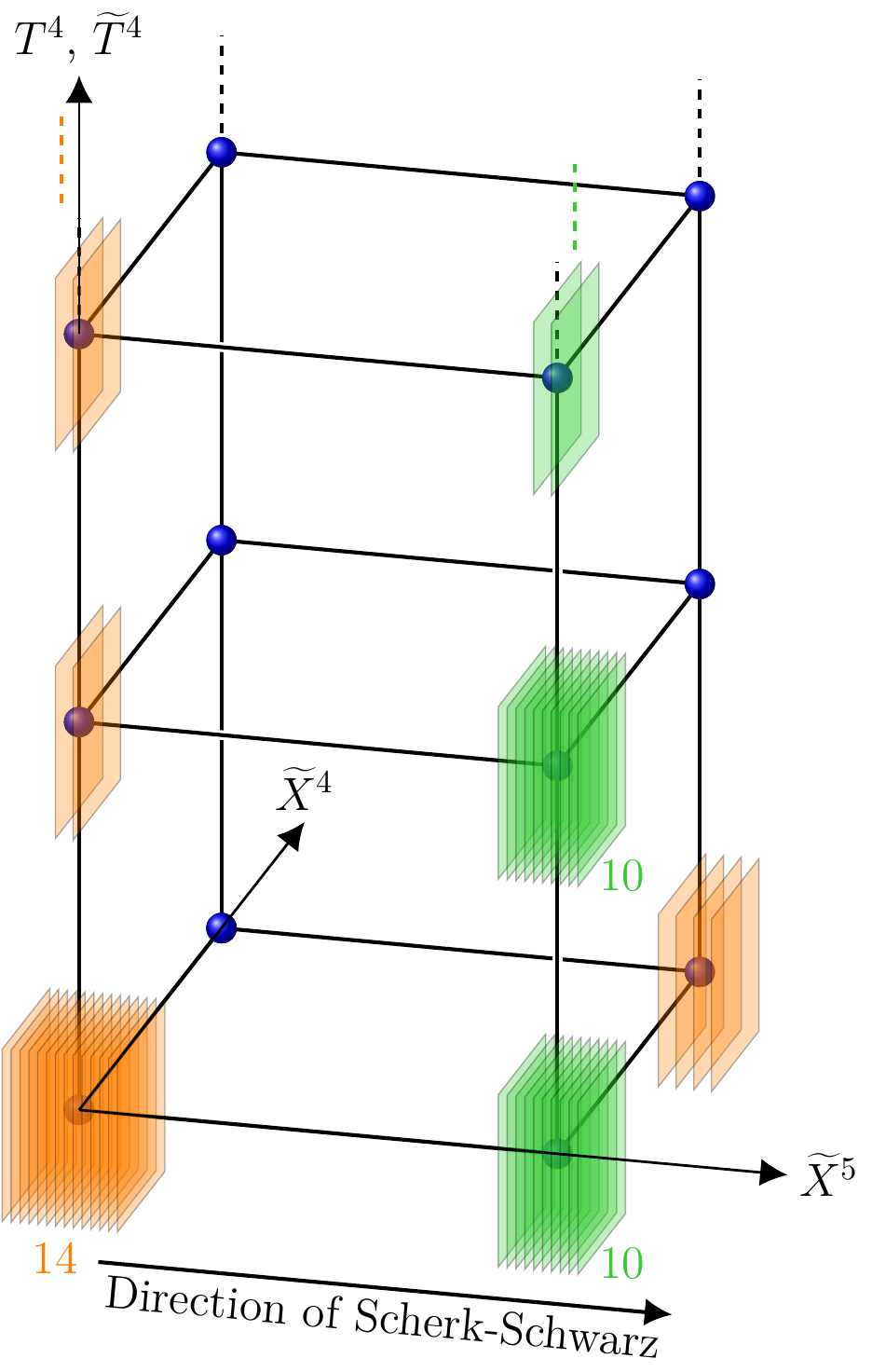}
\end{center}
\caption{\footnotesize Brane configuration tachyon free at one loop, with $\nF=\nB$ and gauge group $[U(1)^7\times U(2)\times U(7)]_{\text{DD}}\times [U(1)^6\times U(5)^2]_{\text{NN}}$.}
\label{nfnb1}
\end{subfigure}
\quad
\begin{subfigure}[b]{0.31\textwidth}
\begin{center}
\includegraphics [scale=0.54]{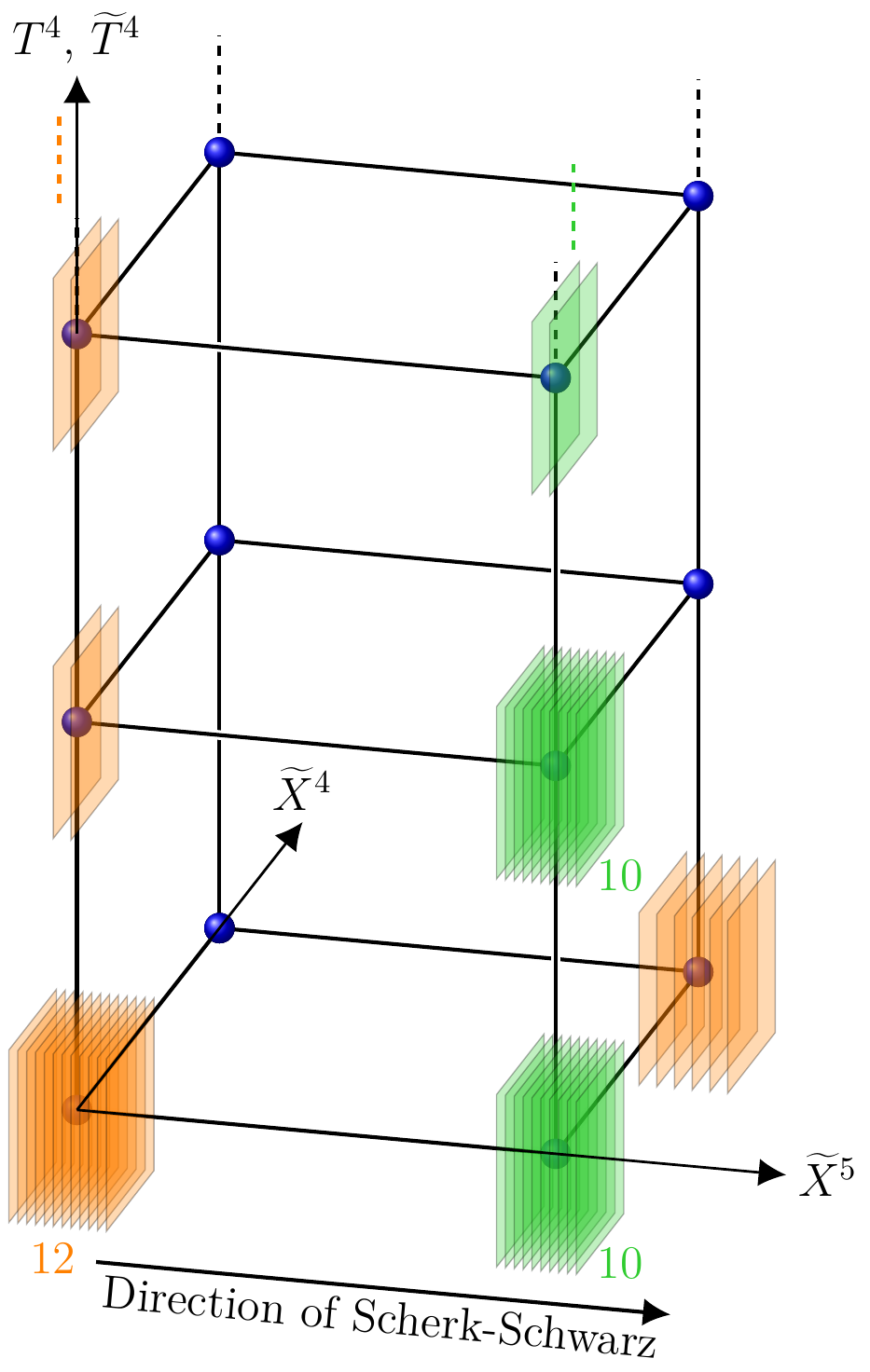}
\end{center}
\caption{\footnotesize Brane configuration tachyon free at one loop, with $\nF=\nB$ and gauge group $[U(1)^7\times U(3)\times U(6)]_{\text{DD}}\times [U(1)^6\times U(5)^2]_{\text{NN}}$.}
\label{nfnb2}
\end{subfigure}
\quad
\begin{subfigure}[b]{0.31\textwidth}
\begin{center}
\includegraphics [scale=0.54]{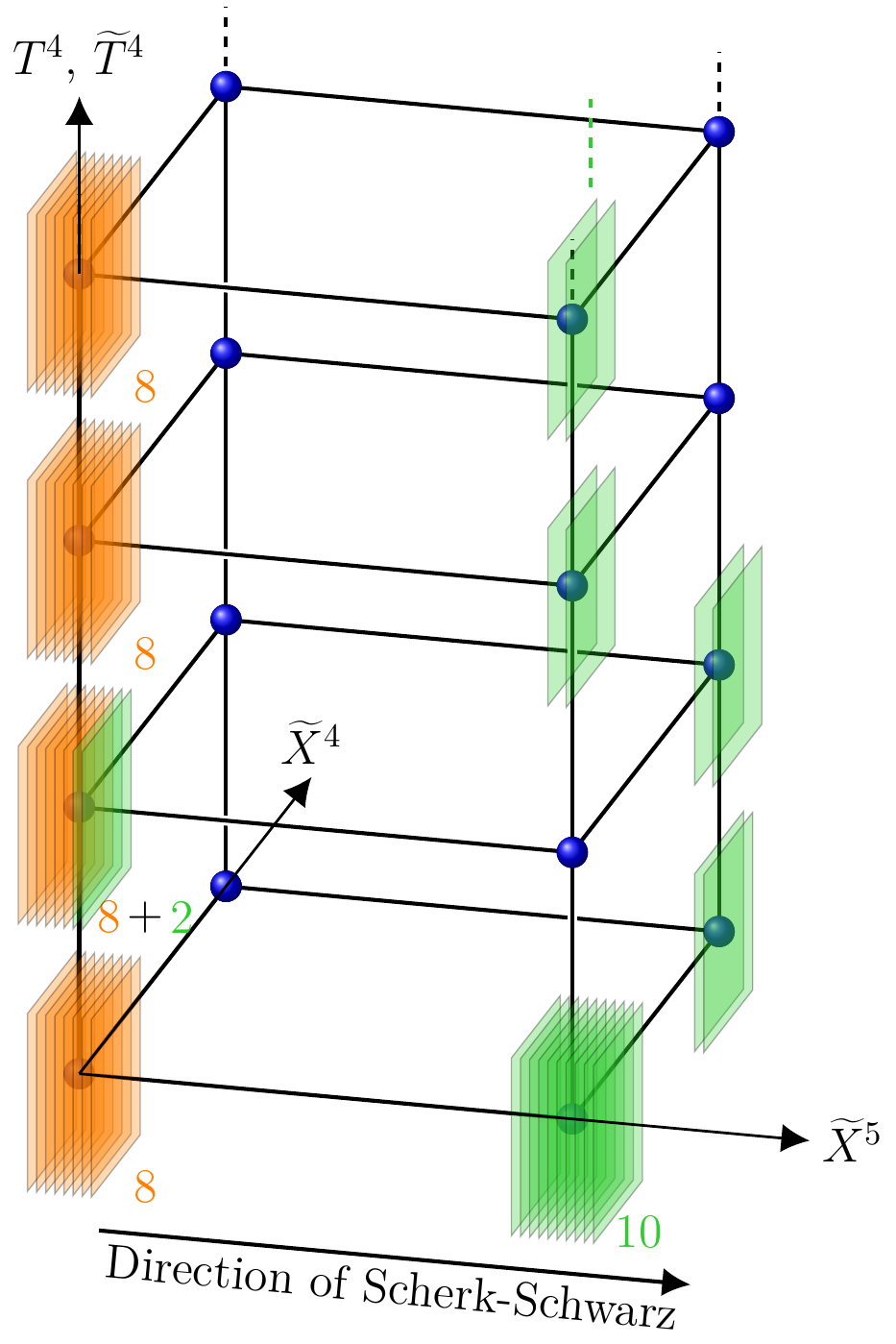}
\end{center}
\caption{\footnotesize Brane configuration with moduli in the  ND sector,  with $\nF-\nB=0$ and gauge group $[U(4)^4]_{\text{DD}} \times [U(1)^{11}\times U(5)]_{\text{NN}}$.}
\label{nfnb3}
\end{subfigure}
\caption{\footnotesize Two brane configurations with $\nF-\nB=0$ in the $(\cR,\tilde \cR)=(8,8)$ component of the moduli space and one in the $(\cR,\tilde \cR)=(0,8)$ component.}
\label{examples_3}
\end{center}
\end{figure}

\paragraph{\em Exponentially suppressed potentials:} There are two tachyon free models satisfying  $\nF-\nB=0$ in the component $(\cR,\tilde \cR)=(8,8)$. Their open string gauge groups are  
\begin{align}
\begin{split}
&[U(1)^7\times U(2)\times U(7)]_{\text{DD}}\times [U(1)^6\times U(5)^2]_{\text{NN}}\\
\text{and}\quad&[U(1)^7\times U(3)\times U(6)]_{\text{DD}}\times U(1)^6\times U(5)^2]_{\text{NN}}~,
\end{split}
\end{align}
and the D3-brane configurations are  depicted in Figs~\ref{nfnb1} and \ref{nfnb2}, respectively. In the first case, the D3-branes T-dual to the D5-branes are distributed in $T^4/\Z_2$ as  7 pairs and one stack of 18 D3-branes, which is split in $\tilde T^2/I_{45}$ into $14+4$ branes.  The D3-branes T-dual to the D9-branes are distributed as 6 pairs and two stacks of 10. The  second configuration is identical to the previous one, up to the splitting of the 18 D3-branes now  into $12+6$.

In both cases, all position moduli along $T^4/\Z_2$ or $\tilde T^4/\Z_2$ are rigid or massive, since condition (\ref{condiwl}) is satisfied. Moreover,  because there are 17 unitary gauge group factors, all of the 16 twisted quaternionic scalars get a mass thanks to the Green-Schwarz mechanism. The latter also  implies all WL's along $\tilde T^2/I_{45}$ to be massive. Moreover, the ND sector does not contain moduli fields. The one-loop potential admits flat directions parametrised by the internal metric (including $G^{55}$ \ie $M$), the dilaton,  and the RR two-form. Notice that these configurations exist in four dimensions but not in five.

A third model with a vanishing potential exists in the component $(\cR,\tilde \cR)=(0,8)$ of the moduli space. Its gauge group is 
\begin{equation}
[U(4)^4]_{\text{DD}} \times [U(1)^{11}\times U(5)]_{\text{NN}}~,
\end{equation}
and the brane configuration is depicted in Fig.~\ref{nfnb3}. The D3-branes T-dual to the D5-branes are distributed in $T^4/\Z_2$ as 4 stacks of 8 D3-branes. The D3-branes T-dual to the D9-branes are distributed as 8 pairs, one stack of 12 branes which is split in $\tilde T^2/I_{45}$ into $10+2$ and one stack of 4 branes which is split in $\tilde T^2/I_{45}$ into $2+2$. In this model, the position moduli along $T^4/\Z_2$ or $\tilde T^4/\Z_2$ are also rigid or massive and all the twisted quaternionic scalars get a tree-level mass. After implementation of the Green-Schwarz mechanism, the WL's along $T^2$ are all massless except one which is massive. We cannot however conclude on the full stability of the model since moduli in the ND sector are present in this configuration.

%%%%%

\begin{figure}[H]
\captionsetup[subfigure]{position=t}
\begin{center}
\begin{subfigure}[t]{0.31\textwidth}
\begin{center}
\includegraphics [scale=0.5]{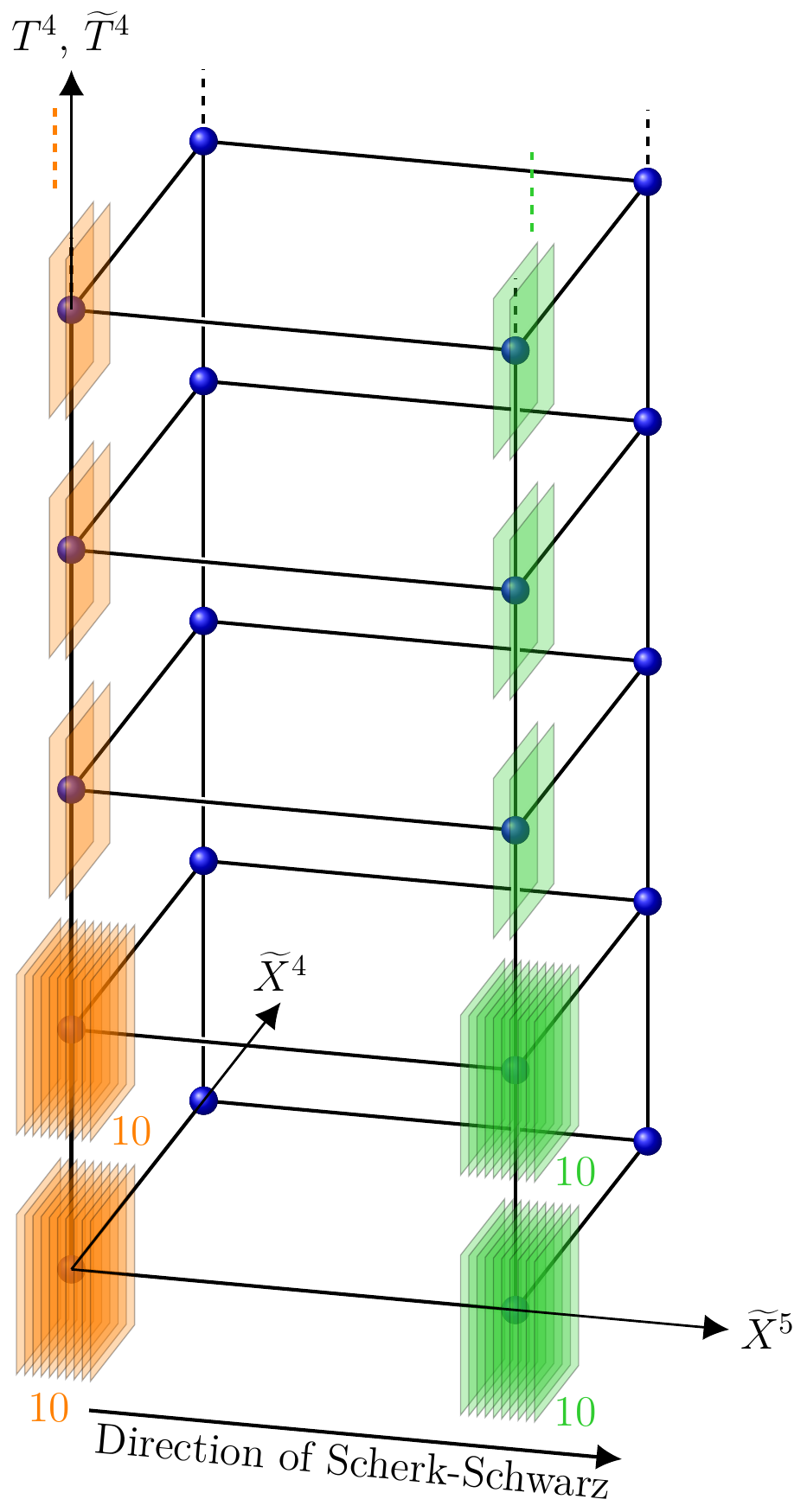}
\end{center}
\caption{\footnotesize Brane configuration tachyon free at one loop, with $\nF-\nB=40$.}
\label{examples_3a}
\end{subfigure}
\quad
\begin{subfigure}[t]{0.31\textwidth}
\begin{center}
\includegraphics [scale=0.5]{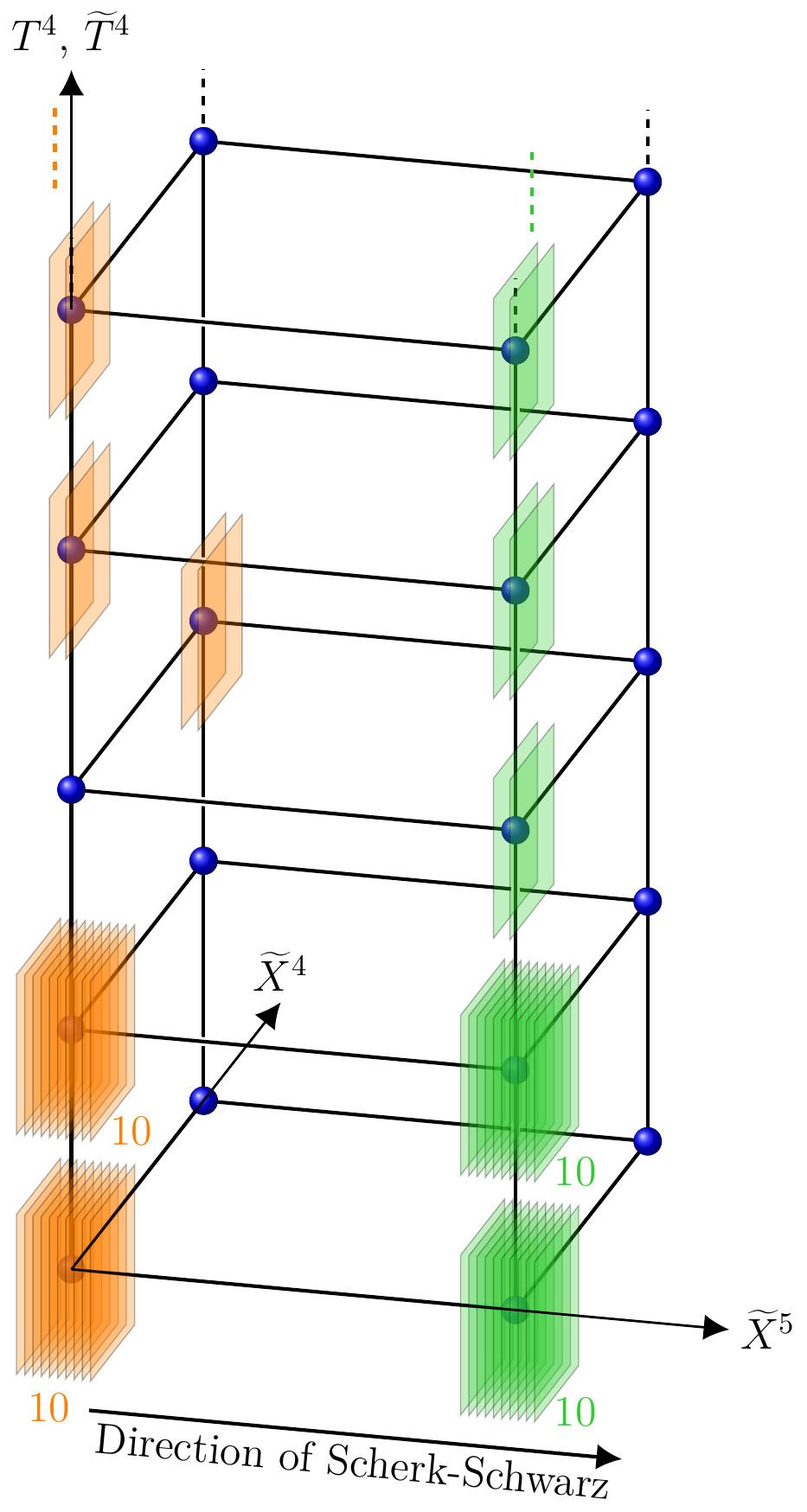}
\end{center}
\caption{\footnotesize Brane configuration tachyon free at one loop, with $\nF-\nB=24$.}
\label{examples_3b}
\end{subfigure}
\quad
\begin{subfigure}[t]{0.31\textwidth}
\begin{center}
\includegraphics [scale=0.5]{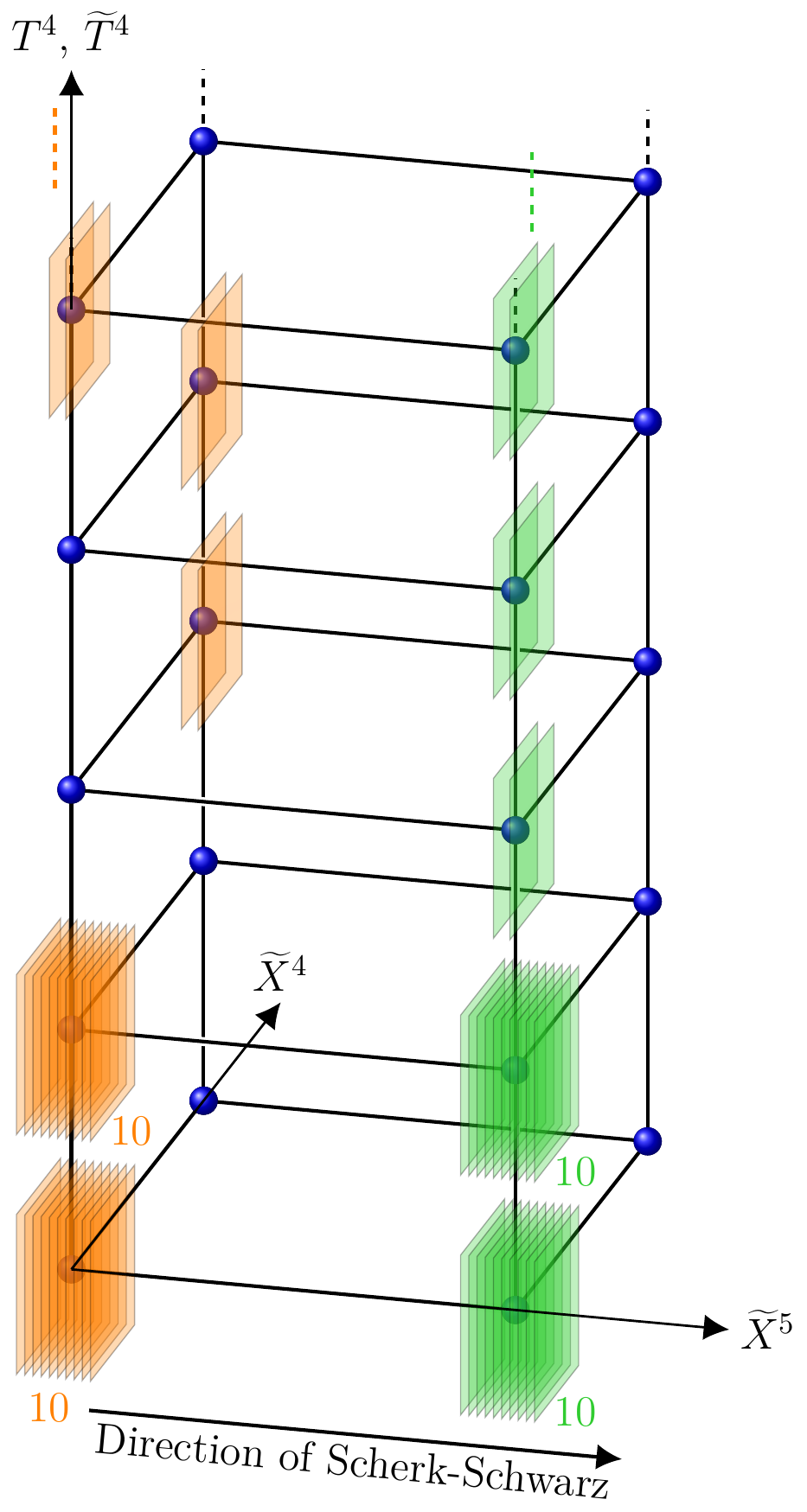}
\end{center}
\caption{\footnotesize Brane configuration tachyon free at one loop, with $\nF-\nB=8$.}
\label{examples_3c}
\end{subfigure}
\quad
\begin{subfigure}[t]{0.31\textwidth}
\begin{center}
\includegraphics [scale=0.5]{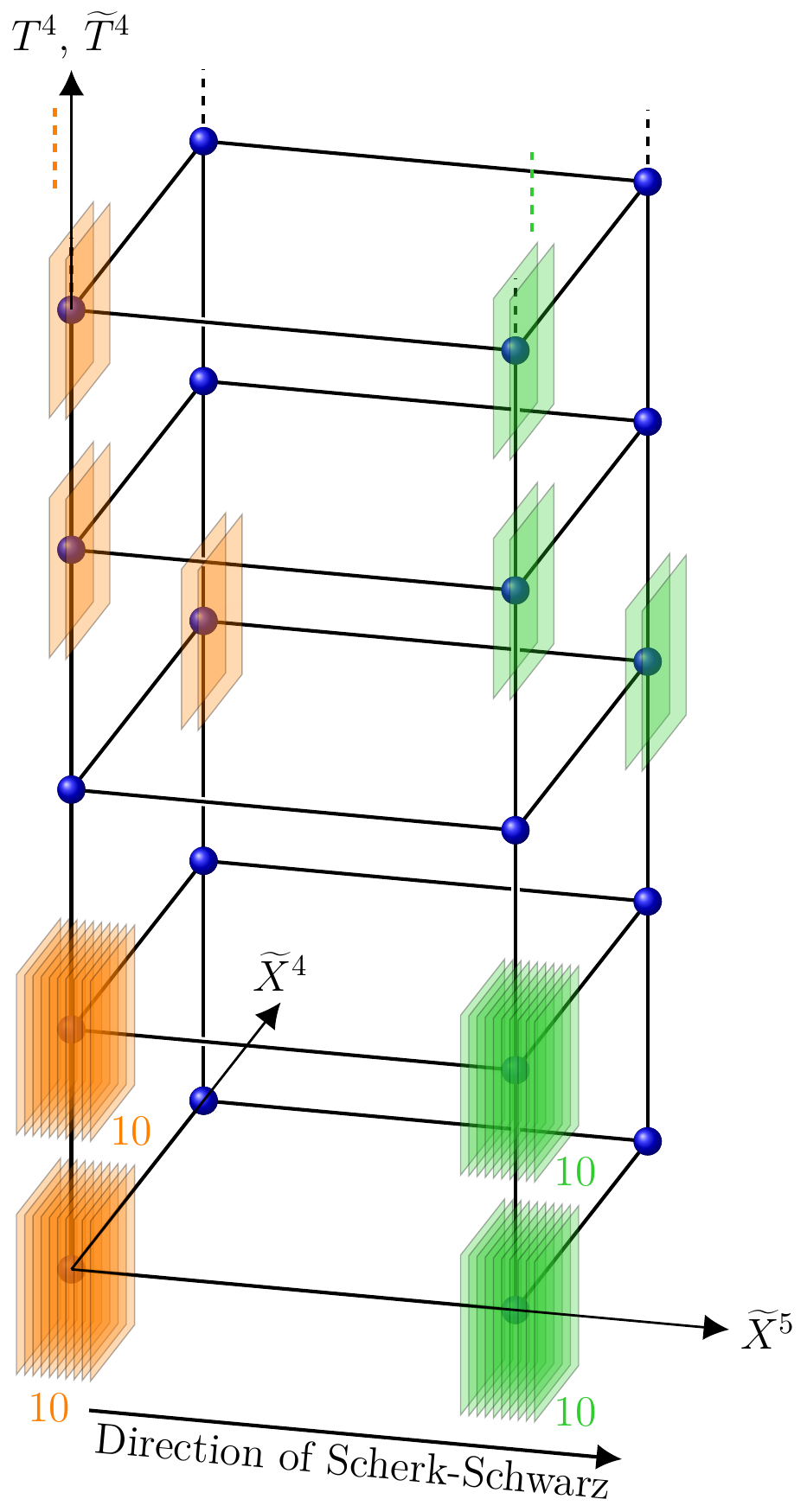}
\end{center}
\caption{\footnotesize Brane configuration tachyon free at one loop, with $\nF-\nB=10$.}
\label{examples_3d}
\end{subfigure}
\quad
\begin{subfigure}[t]{0.31\textwidth}
\begin{center}
\includegraphics [scale=0.5]{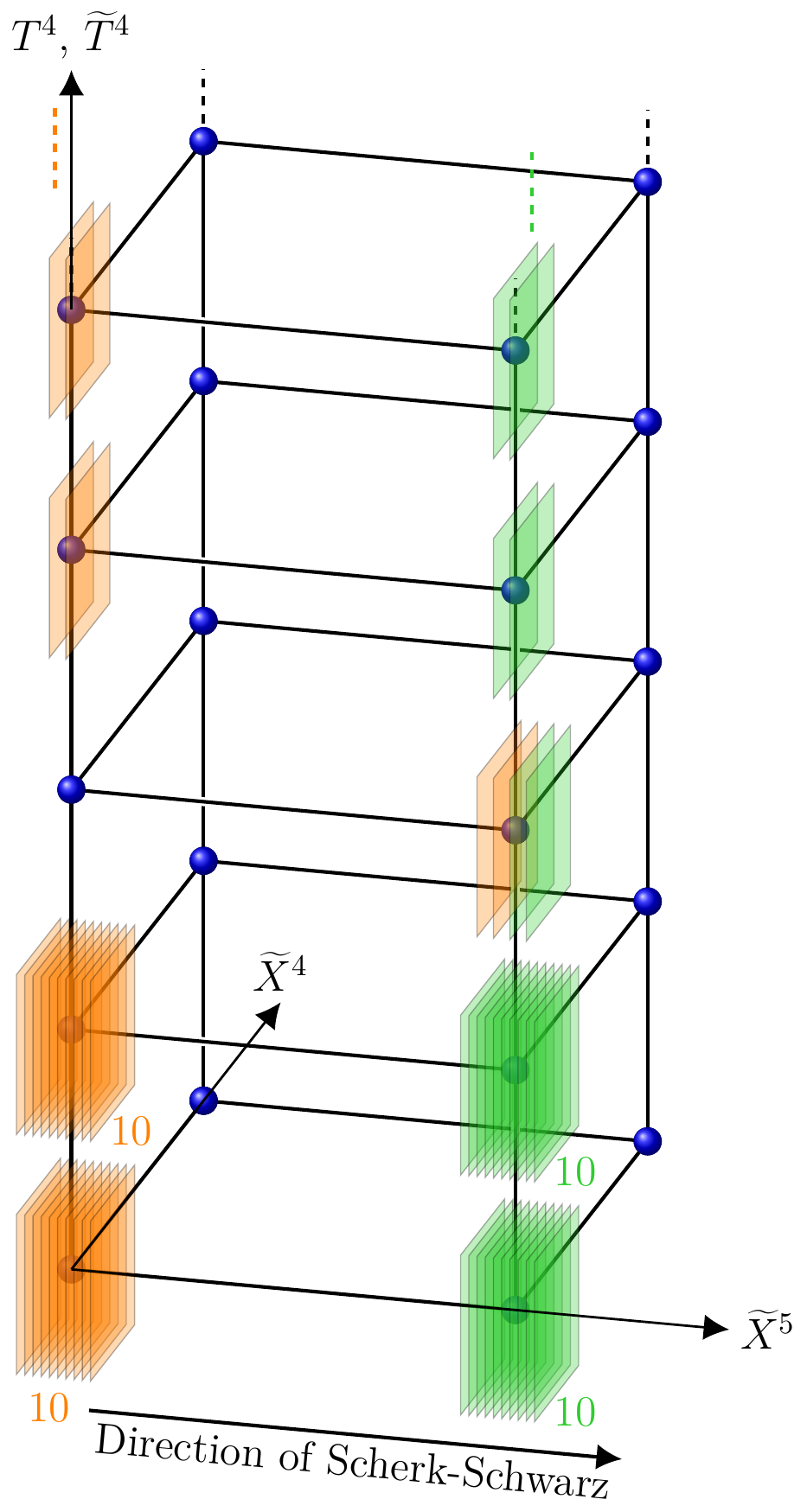}
\end{center}
\caption{\footnotesize Brane configuration with \mbox{$\nF-\nB=8$}.}
\label{examples_3e}
\end{subfigure}
\caption{\footnotesize Brane configurations with $\nF-\nB>0$.}
\label{examples_3}
\end{center}
\end{figure}

\paragraph{\em Positive potentials:} Let us analyze the five configurations shown in Fig.~\ref{examples_3a}--\ref{examples_3e}, 
which yield an identical open string gauge group $[U(1)^6\times U(5)^2]_{\text{DD}}\times[U(1)^6\times U(5)^2]_{\text{NN}}$. All position moduli along $T^4/\Z_2$ and $\tilde T^4/\Z_2$  are rigid or massive, while  the WL's along $\tilde T^2/I_{45}$ are either massive or massless, depending on the case at hand, thanks to  the Green--Schwarz mechanism. There are no moduli in the ND sector except for the last configuration. Moreover, because there are 16 unitary gauge group factors, all moduli belonging to the twisted quaternionic scalars are massive.

The configuration in Fig.~\ref{examples_3a} yields $\nF-\nB=40$. Notice that it may be considered in five dimensions, by decompactifying the direction $X^4$.  In the case shown in Fig.~\ref{examples_3b}, the direction $\tilde X^4$ is used to isolate one pair of D3-branes, which leads to  $\nF-\nB=24$. By  isolating a second pair of the same kind as depicted in Fig.~\ref{examples_3c}, one obtains $\nF-\nB=8$.  Starting back from the configuration with $\nF-\nB=24$, one can obtain $\nF-\nB=10$ by isolating a second pair of the other kind as depicted in Fig.~\ref{examples_3d}. Finally, one may consider the configuration in Fig.~\ref{examples_3e}, which also leads to $\nF-\nB=8$, but contains  quaternionic scalars in the ND sector whose masses need to be analyzed at one loop. 

%%%%%%%%%%%%%%%%%%%%%%%%%%%%%%%%%%%%%

%%%%%%%%%%%%%%%%%%%%%%%%%%%%%%%%%%%%%%


\begin{thebibliography}{99}

\bibitem{Paper}
  S.~Abel, T.~Coudarchet and H.~Partouche,
  ``On the stability of open string orbifold models with broken supersymmetry,''
  arXiv:2003.02545 [hep-th].
  %%CITATION = ARXIV:2003.02545;%%
  
\bibitem{PreviousPaper}
  S.~Abel, E.~Dudas, D.~Lewis and H.~Partouche,
  ``Stability and vacuum energy in open string models with broken supersymmetry,''
  JHEP {\bf 1910} (2019) 226
  %doi:10.1007/JHEP10(2019)226
  [arXiv:1812.09714 [hep-th]].
  
\bibitem{Partouche:2019pgv}
  H.~Partouche,
  ``Quantum stability in open string theory with broken supersymmetry,''
  arXiv:1901.02428 [hep-th].
  
  
%%%%%%%%%%%%%%%%
\bibitem{Itoyama:1986ei}
  H.~Itoyama and T.~R.~Taylor,
  ``Supersymmetry restoration in the compactified $O(16) \times O(16)'$ heterotic string theory,''
  Phys.\ Lett.\ B {\bf 186} (1987) 129.
  
 \bibitem{Abel:2015oxa} 
  S.~Abel, K.~R.~Dienes and E.~Mavroudi,
  ``Towards a non-supersymmetric string phenomenology,''
  Phys.\ Rev.\ D {\bf 91} (2015) 126014
  [arXiv:1502.03087 [hep-th]].
  
  \bibitem{SNS1}
  C.~Kounnas and H.~Partouche,
  ``Super no-scale models in string theory,''
  Nucl.\ Phys.\ B {\bf 913} (2016) 593
  %doi:10.1016/j.nuclphysb.2016.10.001
  [arXiv:1607.01767 [hep-th]].

\bibitem{SNS2}
  C.~Kounnas and H.~Partouche,
  ``$\N=2 \to 0$ super no-scale models and moduli quantum stability,''
  Nucl.\ Phys.\ B {\bf 919} (2017) 41
  %doi:10.1016/j.nuclphysb.2017.03.011
  [arXiv:1701.00545 [hep-th]].
  
\bibitem{FR}
  I.~Florakis and J.~Rizos,
  ``Chiral heterotic strings with positive cosmological constant,''
  Nucl.\ Phys.\ B {\bf 913} (2016) 495
  [arXiv:1608.04582 [hep-th]].
  
  \bibitem{Abel:2017rch}
  S.~Abel and R.~J.~Stewart,
  ``On exponential suppression of the cosmological constant in non-SUSY strings at two loops and beyond,''
  Phys.\ Rev.\ D {\bf 96} (2017) 106013
  [arXiv:1701.06629 [hep-th]].
  

  %\cite{Abel:2017vos}
\bibitem{Abel:2017vos}
 S.~Abel, K.~R.~Dienes and E.~Mavroudi,
  ``GUT precursors and entwined SUSY: The phenomenology of stable nonsupersymmetric strings,''
  Phys.\ Rev.\ D {\bf 97} (2018) no.12,  126017
%  doi:10.1103/PhysRevD.97.126017
  [arXiv:1712.06894 [hep-ph]].
  %%CITATION = ARXIV:1712.06894;%%
  
  \bibitem{CatelinJullien:2007hw}
  T.~Catelin-Jullien, C.~Kounnas, H.~Partouche and N.~Toumbas,
  ``Thermal/quantum effects and induced superstring cosmologies,''
  Nucl.\ Phys.\ B {\bf 797} (2008) 137
  %doi:10.1016/j.nuclphysb.2007.12.030
  [arXiv:0710.3895 [hep-th]].
 
  
  \bibitem{Bourliot:2009na}
  F.~Bourliot, C.~Kounnas and H.~Partouche,
  ``Attraction to a radiation-like era in early superstring cosmologies,''
  Nucl.\ Phys.\ B {\bf 816} (2009) 227
 % doi:10.1016/j.nuclphysb.2009.03.006
  [arXiv:0902.1892 [hep-th]].
  
      \bibitem{CFP}
  T.~Coudarchet, C.~Fleming and H.~Partouche,
  ``Quantum no-scale regimes in string theory,''
  Nucl.\ Phys.\ B {\bf 930} (2018) 235
 % doi:10.1016/j.nuclphysb.2018.03.002
  [arXiv:1711.09122 [hep-th]].
  
  \bibitem{Borunda:2002ra}
  M.~Borunda, M.~Serone and M.~Trapletti,
  ``On the quantum stability of IIB orbifolds and orientifolds with Scherk-Schwarz SUSY breaking,''
  Nucl.\ Phys.\ B {\bf 653} (2003) 85
 % doi:10.1016/S0550-3213(03)00040-3
  [hep-th/0210075].
  %%CITATION = doi:10.1016/S0550-3213(03)00040-3;%%
  %23 citations counted in INSPIRE as of 24 Mar 2020

    \bibitem{APP}
  C.~Angelantonj, H.~Partouche and G.~Pradisi,
  ``Heterotic-type I dual pairs, rigid branes and broken SUSY,''
  Nucl.\ Phys.\ B {\bf 954} (2020) 114976
 % doi:10.1016/j.nuclphysb.2020.114976
  [arXiv:1912.12062 [hep-th]].
  
  
     \bibitem{sta3}
  T.~Catelin-Jullien, C.~Kounnas, H.~Partouche and N.~Toumbas,
  ``Induced superstring cosmologies and moduli stabilization,''
  Nucl.\ Phys.\ B {\bf 820} (2009) 290
  [arXiv:0901.0259 [hep-th]].

  \bibitem{sta1}
  F.~Bourliot, J.~Estes, C.~Kounnas and H.~Partouche,
  ``Cosmological phases of the string thermal effective potential,''
  Nucl.\ Phys.\ B {\bf 830} (2010) 330
%  doi:10.1016/j.nuclphysb.2010.01.004
  [arXiv:0908.1881 [hep-th]].
  
  \bibitem{sta2}
  J.~Estes, C.~Kounnas and H.~Partouche,
  ``Superstring cosmology for $\N_4 = 1 \to 0$ superstring vacua,''
  Fortsch.\ Phys.\  {\bf 59} (2011) 861
  %doi:10.1002/prop.201100040
  [arXiv:1003.0471 [hep-th]].


\bibitem{CoudarchetPartouche}
  T.~Coudarchet and H.~Partouche,
  ``Quantum no-scale regimes and moduli dynamics,''
  Nucl.\ Phys.\ B {\bf 933} (2018) 134
  %doi:10.1016/j.nuclphysb.2018.06.009
  [arXiv:1804.00466 [hep-th]].
  %%CITATION = doi:10.1016/j.nuclphysb.2018.06.009;%%
  %13 citations counted in INSPIRE as of 21 Feb 2020 
  
  \bibitem{Itoyama:2020ifw}
H.~Itoyama and S.~Nakajima,
``Stability, enhanced gauge symmetry and suppressed cosmological constant in 9D heterotic interpolating models,''
[arXiv:2003.11217 [hep-th]].


%%%%%%%%%%%%%%%%  
  
   \bibitem{SS}
  J.~Scherk and J.~H.~Schwarz,
  ``Spontaneous Breaking of Supersymmetry Through Dimensional Reduction,''
  Phys.\ Lett.\  {\bf 82B} (1979) 60.
 % doi:10.1016/0370-2693(79)90425-8
  %%CITATION = doi:10.1016/0370-2693(79)90425-8;%%    
	
	\bibitem{openSS2}
	J.~D.~Blum and K.~R.~Dienes,
	``Strong / weak coupling duality relations for nonsupersymmetric string theories,''
	Nucl.\ Phys.\ B {\bf 516} (1998) 83
	% doi:10.1016/S0550-3213(97)00803-1
	[hep-th/9707160].
	%%CITATION = doi:10.1016/S0550-3213(97)00803-1;%%
	%103 citations counted in INSPIRE as of 05 Nov 2018
	
	\bibitem{openSS3}
I.~Antoniadis, E.~Dudas and A.~Sagnotti,
  ``Supersymmetry breaking, open strings and M-theory,''
  Nucl.\ Phys.\ B {\bf 544} (1999) 469
  %doi:10.1016/S0550-3213(98)00806-2
  [hep-th/9807011].
  %%CITATION = doi:10.1016/S0550-3213(98)00806-2;%%
  
  	\bibitem{openSS4}
I.~Antoniadis, G.~D'Appollonio, E.~Dudas and A.~Sagnotti,
  ``Partial breaking of supersymmetry, open strings and M-theory,''
  Nucl.\ Phys.\ B {\bf 553} (1999) 133
 % doi:10.1016/S0550-3213(99)00232-1
  [hep-th/9812118].
  %%CITATION = doi:10.1016/S0550-3213(99)00232-1;%%
 
  \bibitem{openSS5}
    I.~Antoniadis, G.~D'Appollonio, E.~Dudas and A.~Sagnotti, ``Open descendants of $\Z_2 \times \Z_2$ freely acting orbifolds,''
  Nucl.\ Phys.\ B {\bf 565} (2000) 123
 % doi:10.1016/S0550-3213(99)00616-1
  [hep-th/9907184].
  %%CITATION = doi:10.1016/S0550-3213(99)00616-1;%%
     
\bibitem{openSS6}
A.~L.~Cotrone,
  ``A $\Z_2\times \Z_2$ orientifold with spontaneously broken supersymmetry,''
  Mod.\ Phys.\ Lett.\ A {\bf 14} (1999) 2487
  %doi:10.1142/S0217732399002595
  [hep-th/9909116].
  %%CITATION = doi:10.1142/S0217732399002595;%%
  %37 citations counted in INSPIRE as of 19 Dec 2018
  

  
      %\cite{Partouche:2019vzc}
\bibitem{PV1}
  H.~Partouche and B.~de Vaulchier,
  ``Hagedorn-like transition at high supersymmetry breaking scale,''
  JHEP {\bf 1908} (2019) 155
  [arXiv:1903.09116 [hep-th]].
  %%CITATION = doi:10.1007/JHEP08(2019)155;%%
  
  %\cite{Partouche:2019pgz}
\bibitem{PV2}
  H.~Partouche and B.~de Vaulchier,
  ``Phase transition at high supersymmetry breaking scale in string theory,''
  arXiv:1911.06558 [hep-th].
  %%CITATION = ARXIV:1911.06558;%% 
  
\bibitem{Bianchi-Sagnotti}
M.~Bianchi and A.~Sagnotti, 
``Twist symmetry and open string Wilson lines,''
Nucl.\ Phys.\ B {\bf 361} (1991) 519.
  
\bibitem{GimonPolchinski}
  E.~G.~Gimon and J.~Polchinski,
  ``Consistency conditions for orientifolds and d manifolds,''
  Phys.\ Rev.\ D {\bf 54} (1996) 1667
  [hep-th/9601038].
  
\bibitem{GimonPolchinski2}
  M.~Berkooz, R.~G.~Leigh, J.~Polchinski, J.~H.~Schwarz, N.~Seiberg and E.~Witten,
  ``Anomalies, dualities, and topology of $D=6$ $\N=1$ superstring vacua,''
  Nucl.\ Phys.\ B {\bf 475} (1996) 115
  %doi:10.1016/0550-3213(96)00339-2
  [hep-th/9605184].
  
   \bibitem{wip} Work in progress.  
   
   
   
\bibitem{review-3}
  J.~Polchinski,
  ``Tasi lectures on D-branes,''
  hep-th/9611050.
  %%CITATION = HEP-TH/9611050;%%
  %1254 citations counted in INSPIRE as of 04 Jan 2020

   
   \bibitem{dual0}
  C.~Angelantonj, M.~Bianchi, G.~Pradisi, A.~Sagnotti and Y.~S.~Stanev,
  ``Comments on Gepner models and type I vacua in string theory,''
  Phys.\ Lett.\  B {\bf 387} (1996) 743
  [arXiv:hep-th/9607229].

\bibitem{dual1}
  I.~Antoniadis, C.~Bachas, C.~Fabre, H.~Partouche and T.~R.~Taylor,
  ``Aspects of type~I - type~II - heterotic triality in four dimensions,''
  Nucl.\ Phys.\ B {\bf 489} (1997) 160
 % doi:10.1016/S0550-3213(96)00514-7
  [hep-th/9608012].
  
  \bibitem{dual2}
  I.~Antoniadis, H.~Partouche and T.~R.~Taylor,
  ``Duality of $\N=2$ heterotic type~I compactifications in four dimensions,''
  Nucl.\ Phys.\ B {\bf 499} (1997) 29
 % doi:10.1016/S0550-3213(97)00322-2
  [hep-th/9703076].
  
  \bibitem{dual3}
  I.~Antoniadis, H.~Partouche and T.~R.~Taylor,
  ``Lectures on heterotic type I duality,''
  Nucl.\ Phys.\ Proc.\ Suppl.\  {\bf 61A} (1998) 58
   [Nucl.\ Phys.\ Proc.\ Suppl.\  {\bf 67} (1998) 3]
   [NATO Sci.\ Ser.\ C {\bf 520} (1999) 179]
 % doi:10.1016/S0920-5632(98)00114-5
  [hep-th/9706211].
  
  
 

      
\bibitem{PradisiSagnotti}
  G.~Pradisi and A.~Sagnotti,
  ``Open String Orbifolds,''
  Phys.\ Lett.\ B {\bf 216} (1989) 59.
 

\bibitem{review-1}
C.~Angelantonj and A.~Sagnotti,
``Open strings,''
Phys.\ Rept.\  {\bf 371} (2002) 1 [Erratum-ibid.\  {\bf 376} (2003)
339] [arXiv:hep-th/0204089].

\bibitem{review-2}
E.~Dudas,
``Theory and phenomenology of type~I strings and M-theory,''
Class.\ Quant.\ Grav.\  {\bf 17} (2000) R41 [arXiv:hep-ph/0006190].
 
\end{thebibliography}
\end{document}